\documentclass{arxiv}

\usepackage{graphicx}
\usepackage{newtxtext}
\usepackage{newtxmath}
\usepackage{natbib}
\usepackage{hyperref}
\hypersetup{
    colorlinks = true,
    urlcolor   = blue,
    citecolor  = black,
}

\newcommand{\RomanNumeralCaps}[1]

\title{A meso--micro atmospheric perturbation model for wind farm blockage}

\author{Koen Devesse\aff{1}
  \corresp{\email{koen.devesse@kuleuven.be}},
  Luca Lanzilao\aff{1}
 \and Johan Meyers\aff{1}}

\affiliation{\aff{1}Department of Mechanical Engineering, KU Leuven, Leuven, Belgium}

\begin{document}
\maketitle

\begin{abstract}
As wind farms continue to grow in size, mesoscale effects such as blockage and gravity waves become increasingly important. \citet{Dries_Sensitivity_and_feedback} proposed an atmospheric perturbation model (APM) that can simulate the interaction of wind farms and the atmospheric boundary layer while keeping computational costs low. The model resolves the meso-scale flow, and couples to a wake model to estimate the turbine inflow velocities at the micro-scale. This study presents a new way of coupling the mesoscale APM to a wake model, based on matching the velocity between the models throughout the farm. This method performs well, but requires good estimates of the turbine-level velocity fields by the wake model. Additionally, we investigate the mesoscale effects of a large wind farm, and find that aside from the turbine forces and increased turbulence levels, the dispersive stresses due to subgrid flow heterogeneity also play an important role at the entrance of the farm, and contribute to the global blockage effect. By using the wake model coupling, we can explicitly incorporate these stresses in the model. The resulting APM is validated using 27 LES simulations of a large wind farm under different atmospheric conditions. The APM and LES results are compared on both meso- and turbine-scale, and on turbine power output. The APM captures the overall effects gravity waves have on wind farm power production, and significantly outperforms standard wake models.
\end{abstract}


\section{Introduction}
\label{sec:introduction}
As offshore wind farms become larger, their interaction with the atmosphere becomes important to their operation \citep{bleeg2018wind,PorteAgel2020,Fischereit2021}. Recent large-eddy simulation (LES) studies have identified wind-farm induced gravity waves as contributing to the so-called blockage phenomenon, where the wind is slowed down upstream of the farm, thereby reducing turbine power output \citep{Dries_Boundary_layer_development,Dries_Gravity_waves,Lanzilao2022,Lanzilao2023,lanzilao2023parametric,Maas2022,Maas2023,Stipa2023Tosca}.

Offshore, the atmosphere frequently has the structure of a conventionally neutral boundary layer (CNBL), where a neutral atmospheric boundary layer (ABL) is topped by a capping inversion and a stably stratified free atmosphere \citep{Csanady1974,Smedman1997}. Gravity waves can appear in these conditions, with waves generated by surface topography being extensively studied \citep{klemp1975dynamics,Durran_1990,vosper2004inversion,Smith_2007,Orographic_gravity_wave_drag,sachsperger2015lee}. As first hypothesized by \citet{Smith_2010}, wind farms can also trigger such gravity waves. By slowing down the ABL flow, they push the capping inversion upwards, thereby initiating gravity waves both within the inversion layer and in the free atmosphere aloft. The pressure perturbations of these waves in turn affect the flow in the ABL, leading to blockage upstream of the farm, and pressure gradients throughout it \citep{Dries_Boundary_layer_development}. 

While LES studies are a useful tool to gain insight into the flow physics, their high cost prevents them from being used for wind farm planning and control \citep{Meyers2022}. Conventional wake models, which focus on the downstream effects of individual turbines (e.g. \citet{bastankhah2014new}), cannot account for the complex meso-scale effects of large-scale interactions between wind farms and the ABL \citep{PorteAgel2020,Centurelli2021}. Over the years, several approaches have been developed to estimate the global atmospheric response of large wind farms. For instance, the upper limit of wind farm power production has been studied using so-called ``top-down" models, which estimate the available energy density for infinitely large wind farms \citep{Frandsen1992,Frandsen2006,Emeis2009,Calaf2010,Abkar2013}. This approach has also been extended to include entrance and turbine wake effects \citep{Stevens2015,Luzzatto-Fegiz2018}. However, these models focus on turbulent entrainment, and do not include blockage or atmospheric gravity waves.

Meso-scale climate and weather models can fully capture the atmospheric response to wind farms, and have included various wind farm parametrizations \citep{Fischereit2021}. However, the computational cost of these models is high, and their resolution relatively coarse. They do not model turbine--turbine interactions explicitly, as these occur on smaller scales. Recent research by \citet{Nishino2020} produced a rudimentary way of coupling these large-scale flow models to turbine-level simulations, by ensuring that the two scales predict the same average momentum deficit over the farm. While this has produced good agreement with LES for infinite wind farms \citep{Kirby2022}, it still requires expensive numerical models to estimate the atmospheric response for a given farm \citep{Patel2021}. Very recently, \citet{Kirby2023} developed an analytical approach of estimating the momentum availability for a given farm, but do not yet include the effect of upper atmosphere stratification.

Inexpensive linear models of atmospheric gravity waves have been used in mountain wave research for decades \citep{Orographic_gravity_wave_drag}. As first proposed by \citet{Smith_2010}, these models could predict the effects of wind-farm induced gravity waves as well. \citet{Dries_Sensitivity_and_feedback} built on Smith's idea, and produced an atmospheric perturbation model (APM), so named because the effect of the wind farm is modeled as a linear perturbation on the atmospheric flow. It divides the vertical structure of the ABL into two layers of vertically averaged flow, topped by a capping inversion and the stratified free atmosphere. The flow in the ABL layers is modeled explicitly, while the gravity waves are incorporated as a closure equation for the pressure perturbation \citep{Smith_2010,Devesse2022}. In past studies, the model has been called the \textit{three-layer model} (TLM/3LM), but as more layers can be added \citep{Devesse2022}, we will call it the APM. As a simplified meso-scale model, it is relatively fast, and has shown potential for both wind farm control \citep{Lanzilao2021setpoint} and annual energy production estimates \citep{Dries_Annual_Impact}.

The goal of this paper is to improve the representation of wind farms in the APM. As a highly simplified atmospheric perturbation model, the APM can not resolve turbine-level flows, due to its height-averaged modeling approach and coarse grid resolution. Additionally, it does not explicitly model turbulence. Therefore, when simulating large wind farms, the turbine forces, the increase in turbulent momentum entrainment, and the dispersive stresses from the subgrid flow heterogeneity have to be parametrized. Previous studies with the APM neglected the latter two, and represented the wind farm purely in terms of turbine forcing \citep{Dries_Sensitivity_and_feedback,Devesse2022,Stipa2023MSC}. However, based on the LES results from \citet{lanzilao2023parametric}, we find that the increased turbulent momentum flux and dispersive stresses are important to the mesoscale flow. Therefore, we and develop parametrizations to represent them in the APM.

To get an estimate of the turbine-level flow, which is necessary to calculate the turbine inflow velocities and the dispersive stresses, the APM is coupled to a wake model. \citet{Dries_Sensitivity_and_feedback} took the background velocity, to which the wake model adds the turbine wakes, to be constant and equal to the velocity at some distance upstream of the farm entrance (i.e. 10 times the turbine diameter). This approach, while effective at providing a rough estimate of the blockage effect, is very ad-hoc, and can not take the influence of meso-scale phenomena downstream of the farm entrance into account. This is an important shortcoming, as for instance the favorable pressure gradients associated with gravity waves can increase the power output of downstream turbines \citep{Dries_Boundary_layer_development,Lanzilao2022,Lanzilao2023,lanzilao2023parametric,Maas2022,Maas2023}. Very recently, \citet{Stipa2023MSC} improved on this by calculating a background velocity for the wake model based on the APM pressure estimates. We take a different approach, and present a new wake model coupling method based on matching the velocity between the wake model and the APM. In addition, both the new wake model coupling and APM are validated using 27 simulations of a finite wind farm with varying levels of stratification from the LES data-set developed by \citet{lanzilao2023parametric}.

The new coupling method sets up a background velocity for the wake model that ensures a velocity field that is consistent with the APM output. This results in good agreement for both the velocity fields and the turbine power production, provided that the wake model gives a realistic estimation of the turbine-level flow throughout the farm. When analyzing the LES cases from \citet{lanzilao2023parametric}, we find that this requires the inclusion of a turbine induction model. To this end, we use the model proposed by \citet{Troldborg2017}. We use the wake merging method of \citet{Lanzilao2021merging}, as it allows for varying background velocities throughout the farm.

The remainder of this paper is structured as follows. Section \ref{sec:APM} re-derives and describes the APM. Section \ref{sec:wmc} discusses the wake model used in this work, and develops the new coupling method. Section \ref{sec:validation} performs an a priori validation of the coupling method and APM parametrizations using the LES data-set developed by  \citet{lanzilao2023parametric}, and an a posteriori validation of the complete APM against the same data. Finally, section \ref{sec:conclusions} gives some conclusions and suggestions for further research.

\section{Atmospheric Perturbation Model}
\label{sec:APM}

To model the interaction between wind farms and atmospheric gravity waves, this paper further develops the APM introduced by \citet{Dries_Sensitivity_and_feedback}. It is a reduced-order model, where the ABL is treated as two vertically uniform layers of fluid. In the lower layer, also called the wind farm layer, the effect of the turbine forces is felt directly, while the upper layer consists of the remainder of the ABL. These layers are divided by pliant surfaces, so that there is no mass flux between them. The capping inversion, which separates the upper layer and the free atmosphere, also limits momentum flux \citep{Taylor2008}. The model consists of two continuity and momentum equations, one for each layer, and a closure equation that links the thickening of the ABL to the pressure feedback of the gravity waves in the free atmosphere. This section provides an overview of the model's derivation, and its parametrization of gravity waves, turbulent momentum fluxes, and wind-farm effects.

In their original derivation of the APM, \citet{Dries_Sensitivity_and_feedback} added the wind farm forces to the model after the governing equations where derived. This implicitly introduced a filtering operation, in order to represent the relatively small-scale turbines in the mesoscale model. We will largely follow \citet{Dries_Sensitivity_and_feedback}, but explicitly apply this filtering operation and turbine forcing throughout the derivation, which will result in additional dispersive stresses.

\subsection{Derivation}
\label{subsec:derivation}
In the derivation of the APM, two operations are applied to the momentum and continuity equations: a horizontal filtering to obtain the mesoscale flow, and a height-averaging to reduce the order of the model \citep{Dries_Sensitivity_and_feedback}. The former is done through a Gaussian filter, defined as:
\begin{equation}
	\overline{\phi}(x,y) = \int_0^{L_x}\int_0^{L_y}{G(x-x^{\prime},y-y^{\prime})\phi(x^{\prime},y^{\prime})}\mathrm{d}x^{\prime}\mathrm{d}y^{\prime}, \quad \phi = \overline{\phi} + \phi^{\prime\prime},
	\label{eq:filtering}
\end{equation}
where $L_x \times L_y$ is the domain size, and $G$ is a Gaussian kernel with length $L$ \citep{Dries_Sensitivity_and_feedback}:
\begin{equation}
	G(x,y)=\frac{1}{\upi L^2}\mathrm{exp}{\left(-\frac{x^2+y^2}{L^2}\right)}.
	\label{eq:gaussian_kernel}
\end{equation}
Since the APM is reduces the vertical structure of the ABL to two averaged layers, it can only have a meaningful horizontal resolution larger than the ABL height. Therefore, we follow \citet{Dries_Sensitivity_and_feedback} in taking $L=1\mathrm{km}$.

The flow is then split up into two layers, bounded by the ground and the surfaces $z_1$ and $z_2$, which are defined as:
\begin{align}
	\overline{w}(x,y,z_1)&=\overline{\boldsymbol{u}}_h(x,y,z_1)\bcdot\bnabla_h z_1, \\
	\overline{w}(x,y,z_2)&=\overline{\boldsymbol{u}}_h(x,y,z_2)\bcdot\bnabla_h z_2,
\end{align}
where $\boldsymbol{u}_h$ and $\bnabla_h$ are the two-dimensional horizontal velocity vector and del operator, respectively. Note that $z_1$ and $z_2$ are not the filtered pliant surfaces, but rather the pliant surfaces in the filtered flow, as this greatly simplifies the derivation below. However, this difference is negligible in practice. We now define the height-averaging operator:
\begin{align}
	\left\langle\overline{\phi}\right\rangle_1&=\frac{1}{z_1}\int_{0}^{z_1}\overline{\phi}(z)\mathrm{d}z,\label{eq:height_averaging_1}\\
	\left\langle\overline{\phi}\right\rangle_2&=\frac{1}{z_2-z_1}\int_{z_1}^{z_2}\overline{\phi}(z)\mathrm{d}z,\label{eq:height_averaging_2}
\end{align}
where the subscripts 1 and 2 correspond to the lower and upper layer, respectively. Fluctuations around the height-averaged state are denoted using triple-primes, so that $\overline{\phi} = \left\langle\overline{\phi}\right\rangle_i + \overline{\phi}_i^{\prime\prime\prime}$ in layer $i$. We note the resulting layer thicknesses with $h_1$ and $h_2$.

To obtain the APM governing equations, these operators are applied to the steady-state, Reynolds-averaged, incompressible continuity and momentum equations. In doing so, the vertical velocity and the associated momentum equation drop out, analogous to the derivation of the shallow-flow equations. The end result is a continuity and momentum equation for each of the two layers (see \citealt{Dries_Sensitivity_and_feedback} for a detailed derivation):
\begin{align}
	\bnabla_h\bcdot\left(h_1\left\langle\overline{\boldsymbol{u}}_h\right\rangle_{1}\right)&=0,\label{eq:continuity_1}\\
	\bnabla_h\bcdot\left(h_2\left\langle\overline{\boldsymbol{u}}_h\right\rangle_{2}\right)&=0,\label{eq:continuity_2}
\end{align}
\begin{multline}
	\left\langle\overline{\boldsymbol{u}}_h\right\rangle_{1}\bcdot\bnabla_h\left\langle\overline{\boldsymbol{u}}_h\right\rangle_{1} =
	-\frac{1}{\rho_0}\bnabla_h\left\langle\overline{p}\right\rangle_1 
	+ f_c\mathsfbi{J}\bcdot\left(\boldsymbol{U}_{g,h} - \left\langle\overline{\boldsymbol{u}}_h\right\rangle_{1}\right) 
	+ \bnabla_h\bcdot\left\langle\overline{\boldsymbol{\tau}}_{hh}\right\rangle_1 
	+ \frac{\overline{\boldsymbol{\tau}}_{h3,1}-\overline{\boldsymbol{\tau}}_{h3,0}}{h_1} \\
	+ \frac{\left\langle\overline{\boldsymbol{f}}_{\textit{wf}}\right\rangle_1}{h_1} 
	- \left\langle\bnabla\cdot\left(\boldsymbol{\tau}_{d,h}\right)\right\rangle_1 
	- \frac{1}{h_1}\bnabla_h\cdot\left(h_1\left\langle\overline{\boldsymbol{u}}_{h,1}^{\prime\prime\prime}\overline{\boldsymbol{u}}_{h,1}^{\prime\prime\prime}\right\rangle_1\right)
	+ \frac{\boldsymbol{\mathcal{R}}_1}{h_1},
	\label{eq:momentum_1}
\end{multline}
\begin{multline}
	\left\langle\overline{\boldsymbol{u}}_h\right\rangle_{2}\bcdot\bnabla_h\left\langle\overline{\boldsymbol{u}}_h\right\rangle_{2} =
	-\frac{1}{\rho_0}\bnabla_h\left\langle\overline{p}\right\rangle_2 
	+ f_c\mathsfbi{J}\bcdot\left(\boldsymbol{U}_{g,h} - \left\langle\overline{\boldsymbol{u}}_h\right\rangle_{2}\right) 
	+ \bnabla_h\bcdot\left\langle\overline{\boldsymbol{\tau}}_{hh}\right\rangle_2 
	- \frac{\overline{\boldsymbol{\tau}}_{h3,1}}{h_2} \\
	- \left\langle\bnabla\cdot\left(\boldsymbol{\tau}_{d,h}\right)\right\rangle_2 
	- \frac{1}{h_2}\bnabla_h\cdot\left(h_i\left\langle\overline{\boldsymbol{u}}_{h,2}^{\prime\prime\prime}\overline{\boldsymbol{u}}_{h,2}^{\prime\prime\prime}\right\rangle_2\right)
	+ \frac{\boldsymbol{\mathcal{R}}_2}{h_2},
	\label{eq:momentum_2}
\end{multline}
where the indices 1 and 2 indicate the layer number, and the subscripts $h$ and 3 indicate horizontal and vertical components, respectively. Furthermore, $\rho_0$ is the air density, $p$ is the pressure perturbation, $f_c$ is the Coriolis parameter, $\mathsfbi{J}=\boldsymbol{e}_x\boldsymbol{e}_y-\boldsymbol{e}_y\boldsymbol{e}_x$ is the two-dimensional rotation dyadic (with $\boldsymbol{e}_x$ and $\boldsymbol{e}_y$ the unit vectors in the $x$ and $y$ directions, respectively), $\boldsymbol{U}_{g,h}$ is the horizontal geostrophic wind, and $\overline{\boldsymbol{\tau}}_{hh}$ is the horizontal turbulent momentum flux. Large scale pressure gradients are included through the geostrophic balance as $f_c\mathsfbi{J}\bcdot\boldsymbol{U}_{g,h}$. Additionally, $\boldsymbol{\tau}_{h3,0}$ and $\boldsymbol{\tau}_{h3,1}$ are the vertical turbulent kinetic shear stresses at the ground and between the ABL layers, respectively. The wind farm force, also filtered in the horizontal plane, is denoted with $\overline{\boldsymbol{f}}_{\textit{wf}}$. The two penultimate terms in equations \ref{eq:momentum_1} and \ref{eq:momentum_2} are the height-averaged dispersive stresses and the Taylor-shear dispersion, that appear as the filtering and height-averaging operators are applied to the convective acceleration. \citet{Dries_Sensitivity_and_feedback} found the latter of these to be negligible, but did not include the former, as they did not explicitly apply the filtering operation. The dispersive stresses are given by:
\begin{equation}
	\boldsymbol{\tau}_{d,h} = \overline{\boldsymbol{u}\boldsymbol{u}_h}-\overline{\boldsymbol{u}} \ \overline{\boldsymbol{u}}_h.
\end{equation}
Finally, $\boldsymbol{\mathcal{R}}_1$ and $\boldsymbol{\mathcal{R}}_2$ are negligible residual terms, see \citet{Dries_Sensitivity_and_feedback} for a detailed expression.

\subsection{Parametrizations}
\label{subsec:parametrizations}
Equations \ref{eq:continuity_1}-\ref{eq:momentum_2} provide the governing equations for the APM, with the layer thicknesses and velocities $(\langle\overline{\boldsymbol{u}}_h\rangle_{1},h_1)$ and $(\langle\overline{\boldsymbol{u}}_h\rangle_{2},h_2)$ as state variables. We now go through the remaining terms in the momentum equations, and give closure equations and parametrizations where needed.

\subsubsection{Gravity waves}
The displacement of the capping inversion $\eta_t$ causes gravity waves both within the inversion layer and in the free atmosphere above. These waves then induce a pressure perturbation $p_t$ at the top of the ABL, which is given by \citep{Smith_2010}:
\begin{equation}
	\frac{p_t}{\rho_0}=g^{\prime}\eta_t+\mathcal{F}^{-1}\left(\Phi\right)*\eta_t,
	\label{eq:pressure_inv}
\end{equation}
where $\eta_t=h_1+h_2-H$, with $H$ being the unperturbed ABL height, and $*$ denotes a convolution.
The ABL is assumed to be hydrostatic, so the pressure in both layers is equal to this $p_t$. The first term in the above equation gives the pressure feedback of the inversion waves, which directly scales with the reduced gravity $g^{\prime}=g\Delta\theta/\theta$, where the inversion strength $\Delta\theta$ is the jump in potential temperature $\theta$ across the capping inversion. The second term gives the pressure feedback of waves in the free atmosphere. These are generated as the free atmosphere perceives the displacement of the capping inversion similar to large-scale topographies when flowing over it \citep{Smith_2010}. It is most easily expressed in Fourier components, where for each wavenumber $(k,l)$ the pressure scales with the stratification coefficients $\Phi$. For uniform free atmospheres, these coefficients are given by \citep{Smith_2010}:
\begin{equation}
	\Phi=\frac{\mathrm{i}\left(N^2_g-\Omega^2\right)}{m}
	\label{eq:Phi_basic}
\end{equation}
where $N_g=\sqrt{\frac{g}{\theta}\frac{\mathrm{d}\theta}{\mathrm{d}z}}$ is the Brunt-V\"ais\"al\"a frequency, and $\Omega=-\boldsymbol{U}_{g,h}\bcdot(k,l)$ is the intrinsic frequency of the waves. The vertical wavenumber $m$ is given by the dispersion relation \citep{Adrian_Gill_Atmosphere_Ocean_dynamics}:
\begin{equation}
	m^2=\left(k^2+l^2\right)\left(\frac{N^2_g}{\Omega^2}-1\right)
	\label{eq:m2_basic}
\end{equation}
The sign of $m$ has to be chosen so that the waves are evanescent if $m^2<0$, and satisfy the radiation condition if $m^2>0$. The stratification coefficients $\Phi$ can also represent more realistic upper atmospheres, with changes in stratification, wind speed, and wind direction \citep{Devesse2022}.

\subsubsection{Wind-farm model}
The turbine thrust force of each turbine $k$ is calculated as \citep{Dries_Sensitivity_and_feedback}:
\begin{equation}
	\boldsymbol{f}_k = \frac{1}{2}C_{T,k}\frac{\upi D_k^2}{4}S_k^2\boldsymbol{e}_k,
\end{equation}
where $C_{T,k}$ is the thrust coefficient, $D_k$ is the turbine diameter, $S_k$ is the inflow velocity, and $\boldsymbol{e}_k$ is the turbine direction. In this work, we take this direction to be the same for all turbines, so that $\boldsymbol{e}_k=\boldsymbol{e}_t$, and base it on the unperturbed background velocity at the average turbine hub height $z_h$. To account for wake interactions, the turbine inflow velocities are calculated with an engineering wake model. The coupling to this model has been substantially improved compared to the model of \citet{Dries_Sensitivity_and_feedback}, and is the focus of section \ref{sec:wmc}.

The resulting wind-farm force $\boldsymbol{f}_{\textit{wf}}$ is then the sum of the individual turbines:
\begin{equation}
	\boldsymbol{f}_{\textit{wf}} = \sum_k^{N_t}{\boldsymbol{f}_{k}\delta(x-x_k,y-y_k)}
\end{equation}
where $(x_k,y_k)$ denote the turbine locations. The turbine forces are assumed to be point forces, as the grid spacing is much coarser than the turbine diameter.

\subsubsection{Subgrid terms}
Within the wind farm, there are strong velocity fluctuations on the length scale of the turbine diameters. The terms $\langle\bnabla\cdot(\boldsymbol{\tau}_{d,h})\rangle_i$ and $\bnabla_h\cdot(h_i\langle\overline{\boldsymbol{u}}_{h,i}^{\prime\prime\prime}\overline{\boldsymbol{u}}_{h,i}^{\prime\prime\prime}\rangle_i)/h_i$, with $i$ the layer index, represent the mesoscale influence of these unresolved flow features. The former are the dispersive stresses, which are due to horizontal sub-filter variations, while the latter is the result of the vertical flow structure, and sometimes called the Taylor-shear dispersion. Based on an analysis of LES data from \citet{lanzilao2023parametric}, we find that this dispersion is negligible, as was also found by \citet{Dries_Sensitivity_and_feedback}. In contrast, the dispersive stresses are significant within the wind farm, and have to be included in the lower layer (see section \ref{sec:mba} below). The dispersive stresses can be split up into three components:
\begin{equation}
	\left\langle\bnabla\cdot\left(\boldsymbol{\tau}_{d,h}\right)\right\rangle_1 = 
	\bnabla_h\cdot\left\langle\boldsymbol{\tau}_{d,hh}\right\rangle_{1} 
	+ \frac{\left.\boldsymbol{\tau}_{d,h3}\right\vert_{h_1}}{h_1} 
	+ \frac{\left\langle\boldsymbol{\tau}_{d,hh}\right\rangle_{1}-\left.\boldsymbol{\tau}_{d,hh}\right\vert_{h_1}}{h_1}\cdot\bnabla_h h_1
	.
	\label{eq:disp_stress_total}
\end{equation}
The three terms in the right-hand side represent the average horizontal stresses across the layer, the vertical fluxes at the top of the layer, and the vertical variation of the stresses. The latter two of these are negligible compared to the first. Furthermore, we find that to good approximation, the height-averaged dispersive stress term can be simplified as:
\begin{equation}
	\bnabla_h\cdot\left\langle\boldsymbol{\tau}_{d,hh}\right\rangle_{1}  \approx
	\bnabla_h\cdot\left\langle\overline{\boldsymbol{u}_h^{\prime\prime}\boldsymbol{u}_h^{\prime\prime}}\right\rangle_{1}.\label{eq:disp_stress_approx}
\end{equation}
To evaluate the above equation, the horizontal turbine-level velocity field must be known throughout the farm, which can be done using the wake model. Since the wake-model velocity field is also required for the coupling method developed in section \ref{sec:wmc}, this does not increase the computational cost for the APM. The filter and height-averaging operator can then be applied numerically.

\subsubsection{Turbulent momentum fluxes}
\citet{Dries_Sensitivity_and_feedback} modeled the vertical turbulent shear stresses $\boldsymbol{\tau}_{3,0}$ and $\boldsymbol{\tau}_{3,1}$ with constant friction coefficients, assuming that the momentum flux was aligned with the velocity difference across the pliant surfaces. However, this approach does not account for the increase in turbulence caused by large wind farms. To address this, we add a very simple correction term to the momentum flux between the lower and upper layer:
\begin{align}
	\overline{\boldsymbol{\tau}}_{h3,0}&=C\vert\vert\left\langle\overline{\boldsymbol{u}}_{h}\right\rangle_{1}\vert\vert\left\langle\overline{\boldsymbol{u}}_{h}\right\rangle_{1},\label{eq:shear_stress_1}\\
	\overline{\boldsymbol{\tau}}_{h3,1}&=D\vert\vert\left\langle\overline{\boldsymbol{u}}_{h}\right\rangle_{2}-\left\langle\overline{\boldsymbol{u}}_{h}\right\rangle_{1}\vert\vert\left(\left\langle\overline{\boldsymbol{u}}_{h}\right\rangle_{2}-\left\langle\overline{\boldsymbol{u}}_{h}\right\rangle_{1}\right)+\overline{\Delta\boldsymbol{\tau}}_{\textit{wf}}.\label{eq:shear_stress_2}
\end{align}
The friction coefficients $C$ and $D$ are fitted to the unperturbed atmospheric state in the same way as \citet{Dries_Sensitivity_and_feedback}. The added momentum flux $\Delta\boldsymbol{\tau}_{\textit{wf}}$ is modeled as a constant added value in the same direction $\boldsymbol{e}_t$ as the wind farm forcing. It is assumed to have the same shape as the wind farm, shifted downstream to account for internal boundary layer growth. The added term is scaled with the wind farm force density, as is typically done in top-down models for large wind farms \citep{Frandsen1992,Calaf2010,Abkar2013}. The resulting expression for $\Delta\boldsymbol{\tau}_{\textit{wf}}$ is:
\begin{equation}
	\Delta\boldsymbol{\tau}_{\textit{wf}}(x,y) = a_{\tau}\frac{\frac{1}{2}C_TN_t\frac{\upi D^2}{4}\vert\vert\boldsymbol{U}_1\vert\vert^2}{A_{\textit{wf}}}\Pi\left(\boldsymbol{x}-d_{\tau}D\boldsymbol{e}_1\right)\boldsymbol{e}_t,\label{eq:dtau_wf}
\end{equation}
where $C_T$ and $D$ are the average thrust coefficient and turbine diameter of the farm, $A_{\textit{wf}}$ is the wind farm area, $\boldsymbol{U}_1$ is the unperturbed velocity in the lower layer, and $\Pi$ is the footprint of the farm, equal to one within the farm and zero everywhere else. The coefficients $a_{\tau}$ and $d_{\tau}$ are fitted to the LES results of \citet{lanzilao2023parametric}, and are 0.120 and 27.8, respectively. A description of the fitting procedure can be found in Appendix \ref{app:tau_wf_fit}. We note that this approach is quite rudimentary, and future work may look into the development of a more involved model.

The horizontal momentum fluxes are modeled with a simple eddy viscosity formulation \citep{Dries_Sensitivity_and_feedback}:
\begin{equation}
	\overline{\boldsymbol{\tau}}_{hh,1}=\nu_{t,1}\bnabla_h\left\langle\overline{\boldsymbol{u}}_{h}\right\rangle_{1}, \qquad \overline{\boldsymbol{\tau}}_{hh,2}=\nu_{t,2}\bnabla_h\left\langle\overline{\boldsymbol{u}}_{h}\right\rangle_{2},\label{eq:tau_hor}
\end{equation}
where $\nu_{t,1}$ and $\nu_{t,2}$ are the depth-averaged eddy viscosities. The increased turbulence near the wind farm is not taken into account in this term.

\subsection{Linearized equations}
Following \citet{Dries_Sensitivity_and_feedback} we partly linearize the equations \ref{eq:continuity_1}-\ref{eq:momentum_2} around a uniform background state $\left(\boldsymbol{U}_1,H_1\right)$ and $\left(\boldsymbol{U}_2,H_2\right)$ for small perturbations $\left(\boldsymbol{u}^{\prime}_1,\eta_1\right)$ and $\left(\boldsymbol{u}^{\prime}_2,\eta_2\right)$ caused by the wind farm forcing. However, since wake effects at the microscale cannot be accurately represented by small perturbations, we keep the wind farm parametrization non-linear. This results in:
\begin{equation}
	\boldsymbol{U}_1\bcdot\bnabla_h\eta_1+H_1\bnabla_h\bcdot\boldsymbol{u}^{\prime}_1=0,\label{eq:lin_continuity_1}
\end{equation}
\begin{equation}
	\boldsymbol{U}_2\bcdot\bnabla_h\eta_2+H_2\bnabla_h\bcdot\boldsymbol{u}^{\prime}_2=0,\label{eq:lin_continuity_2}
\end{equation}
\begin{eqnarray}
	\boldsymbol{U}_1\bcdot\bnabla_h\boldsymbol{u}^{\prime}_1 & = & -\frac{1}{\rho_0}\bnabla_h p - f_c\mathsfbi{J}\bcdot\boldsymbol{u}^{\prime}_1 + \nu_{t,1}\nabla_h^2\boldsymbol{u}^{\prime}_1 + \frac{\mathsfbi{D}^{\prime}}{H_1}\bcdot\Delta\boldsymbol{u}' - \frac{\mathsfbi{C}^{\prime}}{H_1}\bcdot\boldsymbol{u}^{\prime}_1 - \frac{\boldsymbol{T}_{h3,1}-\boldsymbol{T}_{h3,0}}{H_1^2}\eta_1 \nonumber \\
	&& + \left(\left\langle\overline{\boldsymbol{f}}_{\textit{wf}}\right\rangle_1+\overline{\Delta\boldsymbol{\tau}}_{\textit{wf}}\right)\left(\frac{1}{H_1}-\frac{\eta_1}{H_1^2}\right) + \bnabla_h\cdot\left\langle\overline{\boldsymbol{u}_h^{\prime\prime}\boldsymbol{u}_h^{\prime\prime}}\right\rangle_{1},\label{eq:lin_momentum_1}
\end{eqnarray}
\begin{equation}
	\boldsymbol{U}_2\bcdot\bnabla_h\boldsymbol{u}^{\prime}_2 = -\frac{1}{\rho_0}\bnabla_h p - f_c\mathsfbi{J}\bcdot\boldsymbol{u}^{\prime}_2 + \nu_{t,2}\nabla_h^2\boldsymbol{u}^{\prime}_2 - \frac{\mathsfbi{D}^{\prime}}{H_2}\bcdot\Delta\boldsymbol{u}^{\prime} + \frac{\boldsymbol{T}_{h3,1}}{H_2^2}\eta_2 - \overline{\Delta\boldsymbol{\tau}}_{\textit{wf}}\left(\frac{1}{H_2}-\frac{\eta_2}{H_2^2}\right),\label{eq:lin_momentum_2}
\end{equation}
where $\Delta\boldsymbol{u}'=\boldsymbol{u}^{\prime}_2-\boldsymbol{u}^{\prime}_1$. The matrices $\mathsfbi{C}^{\prime}$ and $\mathsfbi{D}^{\prime}$ are the Jacobians with respect to the velocity perturbations of the turbulent momentum fluxes at the ground and the ABL layer interface, excluding the wind-farm induced momentum flux, respectively \citep{Dries_Sensitivity_and_feedback}. The vectors $\boldsymbol{T}_{h3,0}$ and $\boldsymbol{T}_{h3,1}$ are the unperturbed momentum fluxes $\boldsymbol{\tau}_{h3,0}$ and $\boldsymbol{\tau}_{h3,0}$. Note that \citet{Dries_Sensitivity_and_feedback} did not include these terms, which appear in the linearized momentum equations as the derivatives of the net momentum fluxes over the layers with respect to the layer thicknesses $h_1$ and $h_2$. Physically, this corresponds to the fluxes entering at the layer boundaries being distributed over thicker or thinner layers of fluid. Similarly, the terms containing the wind-farm forces and momentum flux also include the derivatives of the layer thicknesses.

In order to easily incorporate the pressure feedback of the internal gravity waves, the model is solved using a Fourier-Galerkin spectral method. This allows all the terms in equations \ref{eq:lin_continuity_1}-\ref{eq:lin_momentum_2} besides the wind-farm-related terms $\boldsymbol{f}_{\textit{wf}}$, $\Delta\boldsymbol{\tau}_{\textit{wf}}$, and $\bnabla_h\cdot(\overline{\boldsymbol{u}_h^{\prime\prime}\boldsymbol{u}_h^{\prime\prime}})_{1}$ to decouple per wavenumber. To incorporate the wind-farm-related terms, we use a fixed-point iteration solver with a relaxation factor of 0.7. At each step, the decoupled terms form a simple six-by-six matrix for each wavenumer, which is easily solved.

The inputs to the APM are the vertical profiles of the wind speed $\boldsymbol{U}_0(z)$, the potential temperature $\theta(z)$, the vertical momentum flux $\boldsymbol{\tau}_{h3}(z)$, and the wind farm and turbine specifications. From these, the background states and equation coefficients are determined in the same way as by \citet{Dries_Sensitivity_and_feedback}.

\section{Wake model coupling}
\label{sec:wmc}
This section describes the wake model used in this work, and the new method for coupling it to the APM. To increase readability, we drop the overlines, height-averaging brackets, and $h$ subscripts throughout this section, so that $\langle\overline{\boldsymbol{u}}_h\rangle_{1}$ becomes $\boldsymbol{u}_{1}$.

\subsection{Wake model}
To incorporate the spatially varying effects of gravity waves, the wake model should be able to handle heterogeneous background velocities. We therefore employ the wake merging method of \citet{Lanzilao2021merging}. This superposition method conserves mass and momentum as long as the background flow variations have large length scales, which should be the case for gravity-wave induced perturbations. Each turbine multiplies the flow by a wake modifier, so that every turbine wake has a self-similar behavior with respect to the flow without it. For two-directional flow, \citet{Lanzilao2021merging} gave a recursive formula:
\begin{equation}
	\boldsymbol{u}_k(\boldsymbol{x}) = (\boldsymbol{u}_{k-1}(\boldsymbol{x})\cdot\boldsymbol{e}_{\perp,k})(1-W_k(\boldsymbol{x}))\mathbf{e}_{\perp,k} + (\boldsymbol{u}_{k-1}(\boldsymbol{x})\cdot\boldsymbol{e}_{\parallel,k})\boldsymbol{e}_{\parallel,k} \quad k=1,\ldots,N_t,
	\label{eq:wm_2dir}
\end{equation}
\begin{equation}
\boldsymbol{u}_w(\boldsymbol{x}) = \boldsymbol{u}_{N_t}(\boldsymbol{x}), \qquad
\boldsymbol{u}_0(\boldsymbol{x}) = \boldsymbol{U}_b,
\end{equation}
where $N_t$ is the number of turbines, and $W_k$ is the wake deficit function of turbine $k$, defined along the background flow streamlines. The turbines are ordered upstream to downstream, so that the first turbine applies its wake to the background velocity $\boldsymbol{U}_b$. The unit vectors $\boldsymbol{e}_{\perp,k}$ and $\boldsymbol{e}_{\parallel,k}$ denote the directions perpendicular and parallel to the rotor disk of turbine $k$ (see \citet{Lanzilao2021merging} for details).

For the wake model coupling method developed later, it will be important to have an explicit expression for the final velocity field $\boldsymbol{u}_w$. This can be obtained by re-writing equations \ref{eq:wm_2dir} as a matrix multiplication with $\boldsymbol{u}_{k-1}$:
\begin{equation}
	\boldsymbol{u}_k(\boldsymbol{x}) = \mathsfbi{A}_k(\boldsymbol{x})\cdot\boldsymbol{u}_{k-1}(\boldsymbol{x}) \quad k=1,\ldots,N_t,
\end{equation}
where
\begin{equation}
\mathsfbi{A}_k(\boldsymbol{x}) = (1-W_k(\boldsymbol{x}))(\boldsymbol{e}_{\perp,k}\boldsymbol{e}_{\perp,k}) + (\boldsymbol{e}_{\parallel,k}\boldsymbol{e}_{\parallel,k}).
\end{equation}
The explicit formula for two-directional flow is then:
\begin{equation}
\boldsymbol{u}_w(\boldsymbol{x}) = \prod_{k=1}^{N_t}\mathsfbi{A}_k(\boldsymbol{x})\cdot\boldsymbol{U}_{b}(\boldsymbol{x}),
\label{eq:explicit_2dir}
\end{equation}
For simplicity, this paper neglects multi-directional effects and assumes the streamlines are straight throughout the farm along the direction of the background flow and the wind-farm force $\boldsymbol{e}_t$. Throughout this section, all velocities will refer to the velocity components in this direction, unless stated otherwise. An extension of the coupling method to two-directional flow is possible using equation \ref{eq:explicit_2dir}, but beyond the scope of this work. For a given background velocity $U_b$, the wake model then predicts the following velocity field $u_{w}$ \citep{Lanzilao2021merging}:
\begin{equation}
	u_{w}(x,y,z)=U_b(x,y,z)\prod_{k=1}^{N_t}\left[1-W_k(x,y,z)\right].
	\label{eq:wm_velocity}
\end{equation}
For the wake deficit function, we use the Gaussian wake model of \citet{bastankhah2014new}. The evolution of the turbulence intensity is incorporated using the model of \citet{niayifar2016analytical}. The wake model is not tuned, and instead uses the parameters given in the above papers, as these have been found to perform well when compared to operational data \citep{Doekemeijer2022}. The turbines are mirrored to account for ground effects. 

Additionally, an induction zone model is included to accurately represent the velocity field upstream of the turbine. In this work, we use the model by \citet{Troldborg2017} with the parameters reported in \citet{Branlard2020}. The induction model is only used to better represent the velocity field for the coupling method, and is not used when calculating the turbine inflow velocities for the calculation of thrust and power. The necessity of the induction model will be discussed in-depth in section \ref{subsec:nbl_wm_performance}.

\subsection{Velocity matching}
\label{sec:velocity_matching}
The goal of the wake model coupling is to find a background velocity $U_b$ based on a mesoscale APM state. This is done by ensuring that the velocity fields predicted by the wake model and the APM are consistent with each other. Concretely, height-averaging and filtering the wake model velocity as in equations \ref{eq:height_averaging_1} and \ref{eq:filtering} should result in the APM's lower layer velocity field. For a given mesoscale state, the goal is thus to find a heterogeneous background velocity that, once wakes are superimposed on it, matches the mesoscale velocity. This requires the following equation to hold:
\begin{equation}
	\frac{1}{z_1}\int_{0}^{z_1}{\int_0^{L_x}\int_0^{L_y}{G(x-x^{\prime},y-y^{\prime})u_{w}(x^{\prime},y^{\prime},z)}\mathrm{d}x^{\prime}\mathrm{d}y^{\prime}}\mathrm{d}z = u_{1}(x,y),
	\label{eq:velocity_matching}
\end{equation}
where $u_{1}$ is the mesoscale velocity in the lower layer along $\boldsymbol{e}_t$. If the background velocity $U_b$ results in a wake model velocity that satisfies equation \ref{eq:velocity_matching}, the velocity fields of the APM and the wake model are consistent. Because of this concept, we refer to the coupling method developed here as Velocity Matching (VM).

Equation \ref{eq:velocity_matching} matches the velocity fields over the entire computational domain. However, it is unnecessary and computationally costly to obtain the background velocity in regions far away from the farm. We therefore split up the velocity into the parts in- and outside the wind farm:
\begin{equation}
	u_{w}(x,y,z) \approx U_b(x,y,z)\prod_{k=1}^{N_t}\left[1-W_k(x,y,z)\right]\delta_{\textit{wf}}(x,y) + u_{1}(x,y,z)\left(1-\delta_{\textit{wf}}(x,y)\right),
	\label{eq:farm_split}
\end{equation}
where $\delta_{\textit{wf}}=1$ in the region $\Omega_{\textit{wf}}$, which encompasses the wind farm and a buffer region around it with a width of $2L_f$, and $\delta_{\textit{wf}}=0$ everywhere else. There, we assume it is roughly equal to the mesoscale velocity, which should be a good approximation far away from the farm.
The equation for the background velocity then becomes:
\begin{eqnarray}
	&\frac{1}{z_1}\int_{0}^{z_1}{\iint_{\Omega_{\textit{wf}}}{G(x-x^{\prime},y-y^{\prime})U_b(x^{\prime},y^{\prime},z)\prod_{k=1}^{N_t}\left[1-W_k(x^{\prime},y^{\prime},z)\right]}\mathrm{d}x^{\prime}dy^{\prime}}\mathrm{d}z \nonumber\\
	&\quad=u_{1}(x,y)-\frac{1}{z_1}\int_{0}^{z_1}{\int_0^{L_x}\int_0^{L_y}{G(x-x^{\prime},y-y^{\prime})u_{1}(x^{\prime},y^{\prime})\left(1-\delta_{\textit{wf}}(x^{\prime},y^{\prime})\right)}\mathrm{d}x^{\prime}\mathrm{d}y^{\prime}}\mathrm{d}z.
\end{eqnarray}
Finally, to avoid having to solve for the three-dimensional background velocity, we decompose $U_b$ into an unperturbed state $U_{0}(z)$, which is known as an input to the APM, and a perturbation:
\begin{equation}
	U_b(x,y,z) = U_{0}(z) + u_b(x,y)f(z),
\end{equation}
where we use a standard logarithmic shape function for $f$:
\begin{equation}
	f(z) = \frac{1}{\kappa}\log{\left(\frac{z}{z_0}\right)}.
\end{equation}
The resulting final VM equation for the background velocity scale $u_b$ is then:
\begin{align}
	&\frac{1}{z_1}\iint_{\Omega_{\textit{wf}}}G(x-x^{\prime},y-y^{\prime})u_b(x^{\prime},y^{\prime})f(z)\prod_{k=1}^{N_t}\left[1-W_k(x^{\prime},y^{\prime},z)\right]\mathrm{d}x^{\prime}\mathrm{d}y^{\prime}\mathrm{d}z \nonumber\\
	&\quad= u_{1}(x,y) \nonumber\\
	&\quad-\int_0^{L_x}\int_0^{L_y}{G(x-x^{\prime},y-y^{\prime})u_{1}(x^{\prime},y^{\prime},z)\left(1-\delta_{wf}(x^{\prime},y^{\prime})\right)}\mathrm{d}x^{\prime}\mathrm{d}y^{\prime} \nonumber\\
	&\quad-\frac{1}{z_1}\int_{0}^{z_1}{\iint_{\Omega_{\textit{wf}}}{G(x-x^{\prime},y-y^{\prime})U_0(z)\prod_{k=1}^{N_t}\left[1-W_k(x^{\prime},y^{\prime},z)\right]}\mathrm{d}x^{\prime}\mathrm{d}y^{\prime}}\mathrm{d}z.
	\label{eq:velocity_matching_final}
\end{align}
Equation \ref{eq:velocity_matching_final} is a linear equation for the background velocity, which, once solved for, can be used as an input for the wake model. As this essentially requires inverting a filtering operation, the problem is ill-posed. We avoid this issue by phrasing it as a least-squares problem, and limiting the number of degrees of freedom of the background velocity. Equation \ref{eq:velocity_matching_final} is discretized using a variational approach with local first-order shape functions, evenly spaced throughout $\Omega_{\textit{wf}}$. As $u_{1}$ is known at the APM gridpoints, we use a collocation method with the collocation points being the APM gridpoints inside $\Omega_{\textit{wf}}$. Based on a numerical analysis, we choose a ratio of 0.4 shape functions per gridpoint in each direction.

Whether the Gaussian filter or the height-averaging operation is applied first in equation \ref{eq:velocity_matching_final} does not have a large effect. In solving the equation, the height-averaging operator is applied first, as this is faster to compute. The integrals are all computed numerically using numpy and scipy routines, sped-up with the numba package \citep{numba}.

\section{LES-based validation}
\label{sec:validation}
This section contains a validation study of the VM method and the APM based on LES data. Sections \ref{sec:vm_validation} and \ref{sec:mba} contain a priori validations of the coupling method and the parametrizations of the various APM terms, respectively. Section \ref{sec:APM_validation} performs an a posteriori validation campaign of the full APM. 

To validate the APM and the VM coupling method, we use 28 LES cases from \citet{lanzilao2023parametric}. These are all simulations of the same large wind farm of $N_t=160$ IEA 10~MW turbines \citep{bortolotti2019iea}, arranged in a staggered layout with 16~rows and 10~columns. An overview of the atmospheric states and the wind farm can be found in tables \ref{tab:flow_cases} and \ref{tab:wind_farm}, respectively. The original dataset consists of 40 simulations with varying boundary layer heights, capping inversion strengths, and atmospheric stratification levels. We do not use the cases where $H=150\mathrm{m}$, as in those cases the turbine rotors are very close to the capping inversion. We also do not use the neutral stratification simulations, as they have no capping inversion to limit internal boundary layer growth into the free atmosphere. Taking these cases into account would require major modifications of the turbulence parametrization of the APM, which goes beyond the scope of this paper. This leaves 27 cases that will be used throughout the rest of this work to validate the APM. Additionally, we use the neutral stratification simulation with $H=500\mathrm{m}$ for the a priori validation of the VM method in section \ref{sec:vm_validation}. For each of the boundary layer heights, there is also a single turbine simulation, which is used to normalize the power output of the LES results. We follow the naming convention of \citet{lanzilao2023parametric}, where cases are characterized by the boundary layer height, capping inversion strength, and atmospheric lapse rate. For instance, the case with $H=500\mathrm{m}$, $\Delta\theta=5\mathrm{K}$, and $\Gamma=4\mathrm{K/km}$ is denoted as H500-$\Delta\theta$5-$\Gamma$4. Note that these values correspond to the initial conditions, and the exact profiles have changed slightly during the precursor spin-up (see \citet{lanzilao2023parametric} for details).

\begin{table}
  \begin{center}
	\def~{\hphantom{0}}
	  \begin{tabular}{lc}
		       Parameter & Values \\[3pt]
		       $H$ (m) & 300, 500, 1000 \\
		       $\Delta\theta$ (K) & 2, 5, 8 \\
		       $\Gamma$ (K/km) & 1, 4, 8 \\
		       TI (\%) & 4 \\
		       $z_0$ (m) & $1\times10^{-4}$ \\
		       $f_c$ ($\textrm{s}^{-1}$) & $1.14\times10^{-4}$
		  \end{tabular}
	  \caption{Overview of the atmospheric conditions of the flow cases from \citet{lanzilao2023parametric} used for the validation of the coupling method and the APM. In total all 27 combinations of these parameters are considered. Note that these values correspond to the initial profiles, and that the actual conditions changed slightly during the precursor spin-up. See \citet{lanzilao2023parametric} for a detailed description of the simulations.}
	  \label{tab:flow_cases}
	  \end{center}
\end{table}

\begin{table}
\begin{center}
	\def~{\hphantom{0}}
	\begin{tabular}{lc}
		Parameter & Value \\[3pt]
		$N_t$ (-) & 16 \\
		$D$ (m) & 198 \\
		$z_h$ (m) & 119 \\
		$s_x$, $s_y$ (m) & 5D \\
		$C_T$ (-) & 0.88 \\
	\end{tabular}
	\caption{Description of the wind farm from \citet{lanzilao2023parametric} used for the validation of the coupling method and the APM.}
	\label{tab:wind_farm}
\end{center}
\end{table}

To compare the APM on a meso-scale level, and to validate the VM method independently, we construct an APM ground truth from the LES database. In a first step, the velocity and pressure fields are Gaussian-filtered in the horizontal direction with the APM filter length. Then, $z_1$ is obtained by advecting a set of points over the domain starting from height $z=H_1$, to obtain the surface separating the two ABL layers. We initialize these points to coincide with the LES grid along the domain inlet, resulting in 1380 points with a regular spacing of 21.7~m. We find capping inversion displacement by applying the \citet{Rampanelli2004} model, and Gaussian-filtering the results. Finally, the velocities and pressure are height-averaged between the layer boundaries. The pressure at the top of the ABL $p_t$ is evaluated at the bottom of the capping inversion.

\subsection{VM method validation}
\label{sec:vm_validation}
To validate the VM method derived in the previous section, we need APM states for it to couple to. In the current section, we do not yet use the APM for this, but instead use the LES-based states set up above. This allows us to test the coupling's performance separately from the APM.

Section \ref{subsec:nbl_wm_performance} discusses the dependence of the VM method on the performance of the wake model. Section \ref{subsec:vm_performance} validates the coupling using the LES simulations outlined above.

\subsubsection{Importance of wake model performance}
\label{subsec:nbl_wm_performance}
For the VM method to perform well, the wake model needs to provide good estimates for the velocity field. Typically, wake models only need to be accurate near downstream turbines, as the velocity in other regions, such as the near-wake or the induction region, does not affect the inflow velocities for other turbines. As a result, a wake model with an unrealistic near-wake flow field can still perform well, especially when tuned. However, when the VM method couples such a wake model to a mesoscale state, it will try to correct the errors made in the global velocity field, thereby worsening the power prediction. To prevent this, we make three choices when setting up the wake model. First, we mirror the turbines to account for ground effects. Second, we apply a correction to the centerline velocity deficit in the near-wake region \citep{Zong2020}. Third, we include an induction model to account for upstream velocity changes. Here, we use the model by \citet{Troldborg2017}, but other models should also achieve similar performance \citep{Branlard2020}. Note that this induction model is only used to estimate the velocity fields in equation \ref{eq:velocity_matching_final}, and not when performing thrust and power calculations. In the latter it is not necessary, as wake models are designed to yield predictions for these properties.

\begin{figure}
	\centerline{\includegraphics[width=1.00\textwidth]{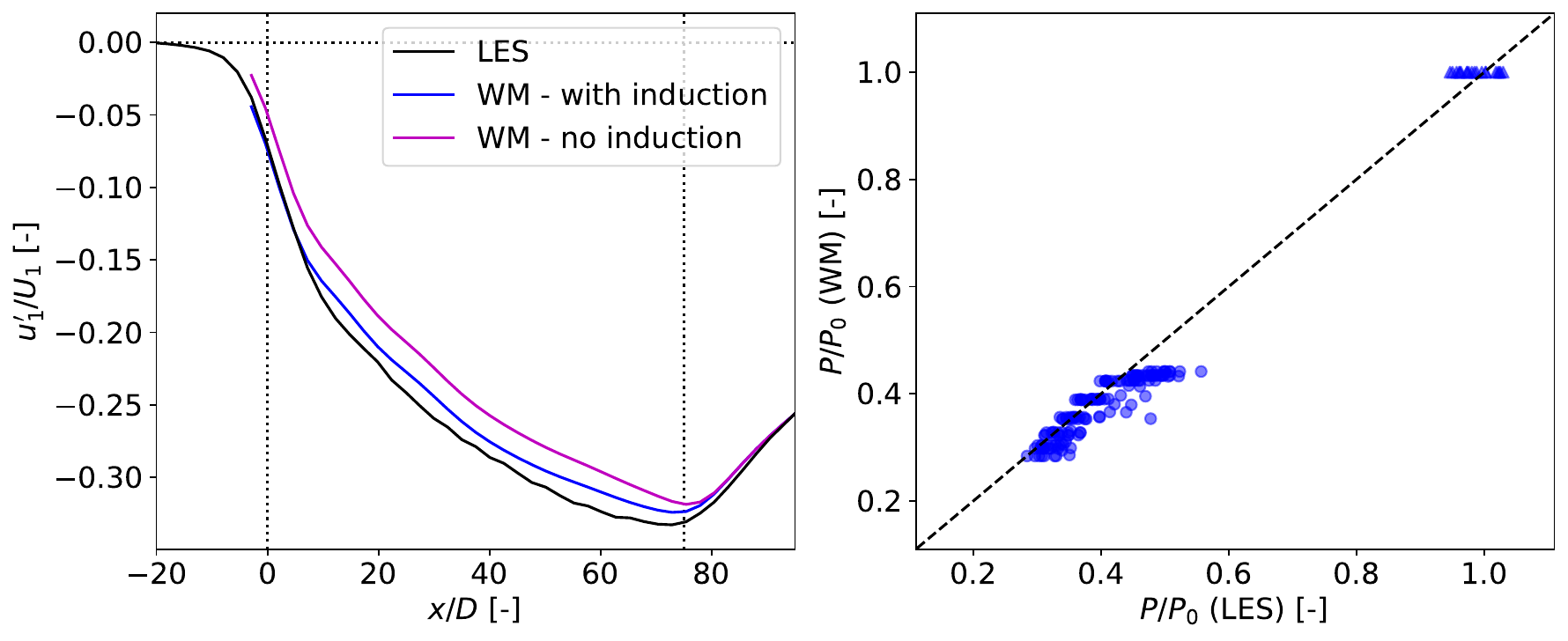}}
	\caption{Performance of the wake model for the case H500-$\Delta\theta$0-$\Gamma$0. Left: the meso-scale velocity deficit as found by LES (black) and the wake model, both with and without induction model (blue and magenta, respectively). The dotted lines denote the wind farm region. Right: the power output as found by LES (x-axis) and wake model (y-axis). The triangles and the circles denote the first two turbine rows and the remainder of the farm, respectively.}
	\label{fig:nbl_wm_performance}
\end{figure}

In order to investigate this dependence on the wake model in detail, we analyze the VM method's performance for the case H500-$\Delta\theta$0-$\Gamma$0. This is a purely neutral case, without gravity waves and minimal mesoscale feedback effects. As a result, one would expect an uncoupled wake model to perform well in terms of both velocity and power, and for the coupling method to predict a background velocity perturbation close to zero. Figure \ref{fig:nbl_wm_performance} shows the mesoscale velocity fields and the power output of the turbines as found by LES and an uncoupled wake model, both with and without induction. The wake model background velocity is simply the precursor profile $U_0(z)$. The LES turbine power outputs were scaled with the power output for a single turbine, and the wake model power outputs are scaled with the power calculation for a single turbine with the precursor as inflow conditions. We observe that the wake model reproduces both the velocity and power accurately, with an average power error across all turbines of 1.8\%. However, without an induction model, the wake model underestimates the velocity upstream of and throughout the farm. 

\begin{figure}
\centerline{\includegraphics[width=1.00\textwidth]{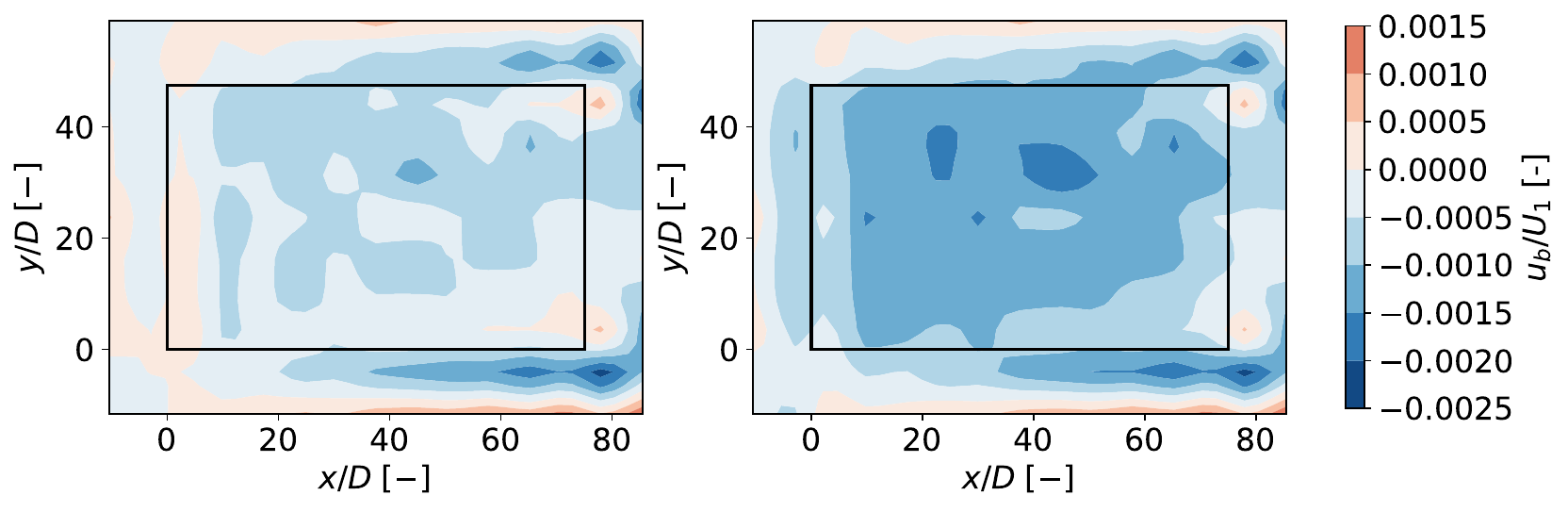}}
\caption{Background velocity perturbations $u_b$ as found by the VM method for the case H500-$\Delta\theta$0-$\Gamma$0 when using a wake model with an induction model (left) and without an induction model (right).}
\label{fig:u_bg_induction}
\end{figure}

We now apply the VM method to this case, matching the wake model velocities to the LES-based APM state. Figure \ref{fig:u_bg_induction} shows the resulting background velocity perturbations $u_b$ with and without using an induction model. When an induction model is used, the VM method predicts almost no variations in the background velocity. Without, the coupling lowers the background velocity considerably. We find that power predictions of the wake model based on this background velocity are up to 10\% too low. 

We conclude that the performance of the wake model drastically affects the VM method's output, and good results depend on the estimations of the turbine-level velocities being realistic at every point in the farm. This requires the wake model to account for turbine induction, the near-wake region, and ground effects. With the corrections used in this work, this is achieved fairly well, although there is still a slight underestimation of the mesoscale velocity deficit within the farm.

\subsubsection{Coupling performance}
\label{subsec:vm_performance}
\begin{figure}
  \centerline{\includegraphics[width=1.00\textwidth]{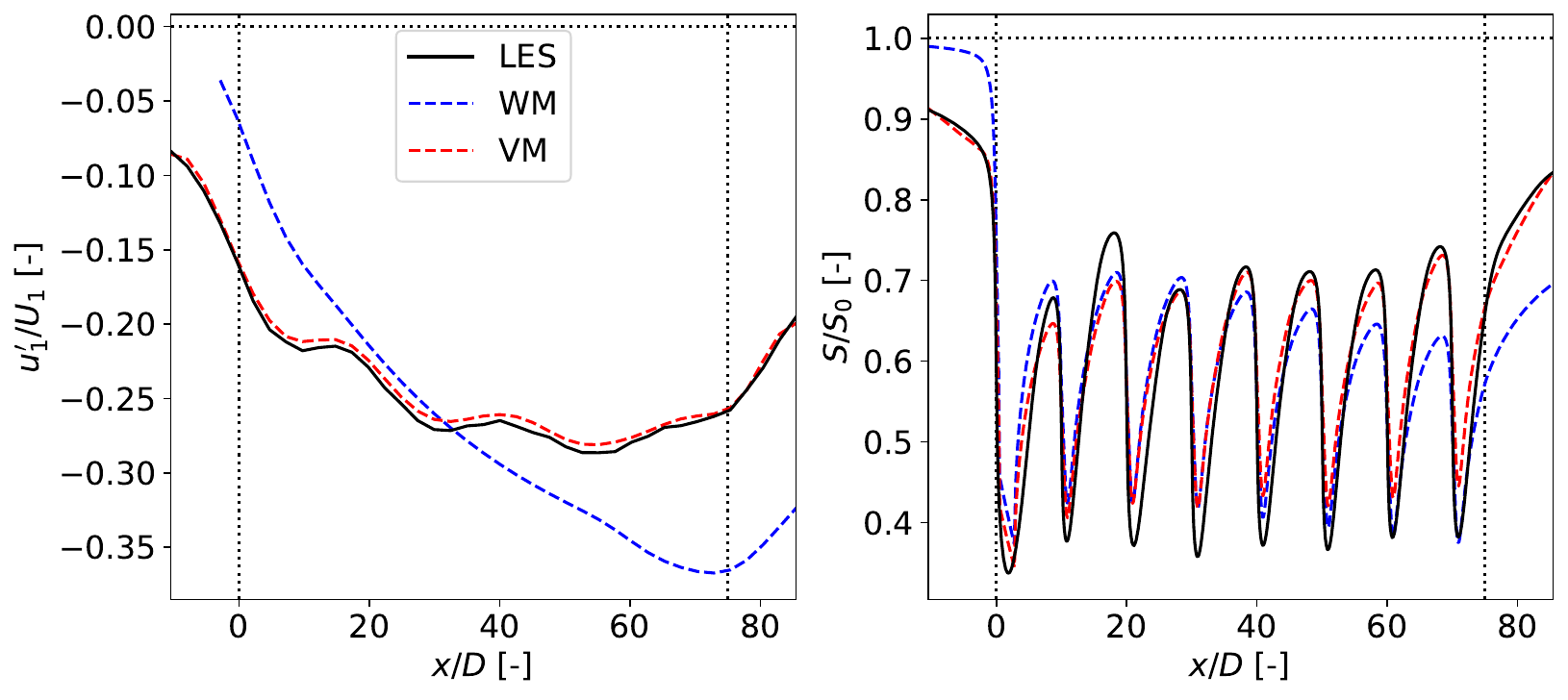}}
  \caption{Streamwise velocity deficits through the center of the farm of the LES (full lines, black) and the wake model (dashed lines), both uncoupled and coupled (blue and red, respectively), for the case H500-$\Delta\theta$5-$\Gamma$4. Left: the meso-scale velocity deficit. Right: the average velocity across a tube with the turbine diameter. The dotted lines denote the wind farm region.}
\label{fig:wmc_velocities}
\end{figure}

We coupled the wake model to the LES-based APM state for all 27 simulations with atmospheric stratification. As a comparison, we also apply the uncoupled wake model to all analyzed cases. The coupling method was consistently able to provide background velocities that resulted in matching mesoscale velocity fields. Figure \ref{fig:wmc_velocities} shows this for the case H500-$\Delta\theta$5-$\Gamma$4. It's clear that this required significant corrections to the background velocity of the wake model, as the uncoupled wake model has a very different profile. Moreover, this mesoscale matching corresponds to a better agreement of the velocity fields on turbine-level. The right side of figure \ref{fig:wmc_velocities} shows the local velocity averaged over a streamwise tube with the turbine diameter, placed at hub height through the center of the farm, for the same case. The velocity-matched wake model captures the lower velocity at the farm entrance well, and follows the LES state throughout the rest of the farm. This good performance is consistent across all tested cases.

\begin{figure}
	\centerline{\includegraphics[width=1.\textwidth]{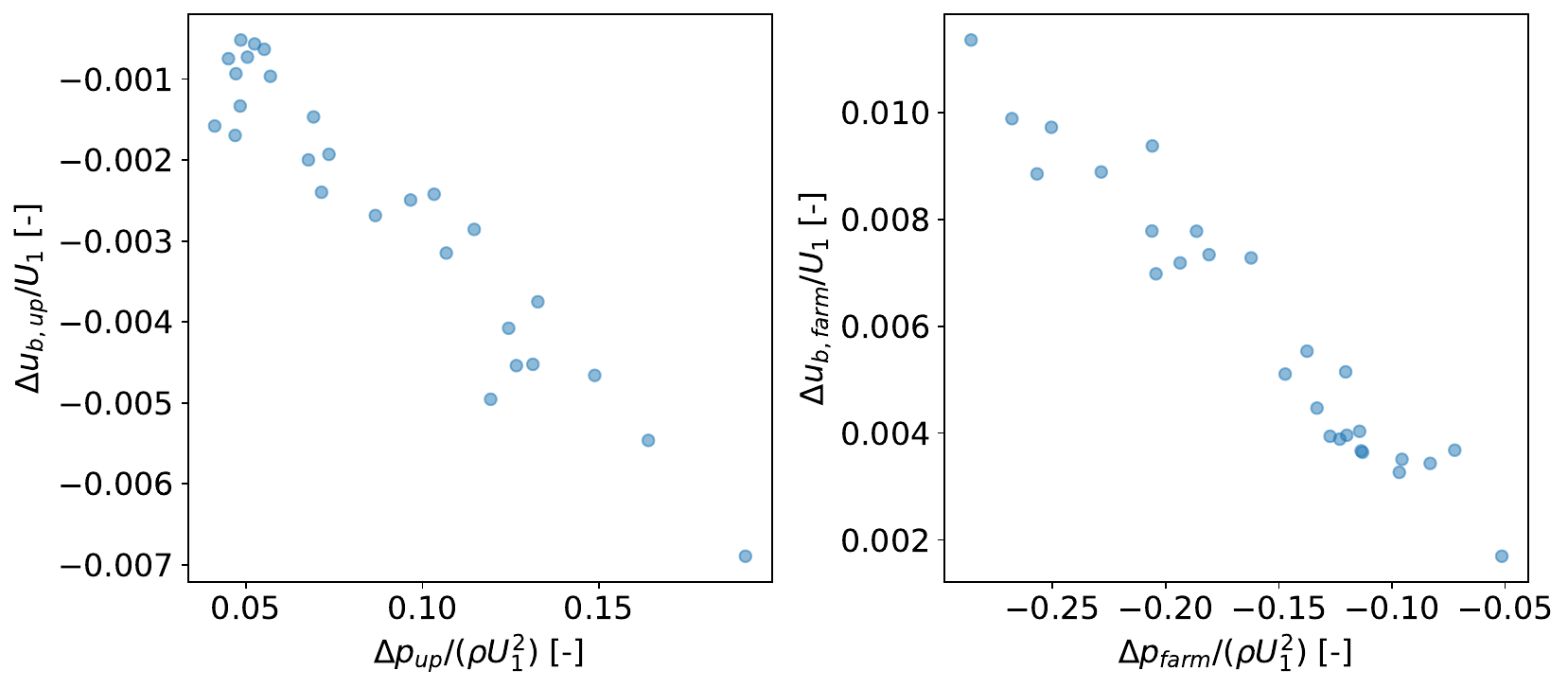}}
	\caption{The averaged pressure (horizontal axes) and background velocity perturbations (vertical axes) over the width of the farm, evaluated at the farm entrance (left) and their difference across the farm (right).}
	\label{fig:ub_dp}
\end{figure}

For all cases, there is a blockage effect, resulting in a lower background velocity at the start of the farm. Throughout the farm itself, the background velocities rise, and even become positive halfway through the farm. This is consistent with the pressure gradients, which \citet{lanzilao2023parametric} found to be unfavorable upstream of the farm, and favorable throughout. As figure \ref{fig:ub_dp} shows, there is a strong correlation between the changes in background velocity and the pressure perturbations, both at the farm entrance and across the farm. Figure \ref{fig:ub_dp} (left) shows the upstream unfavorable pressure rise, which is very well correlated with the upstream change of background velocity. Figure \ref{fig:ub_dp} (right) shows the favorable pressure drop in the farm, which correlates very well with the background velocity change in the farm. Thus the VM method manages to catch the expected favorable and unfavorable pressure gradients through $u_b$.

\begin{figure}
	\centerline{\includegraphics[width=1.00\textwidth]{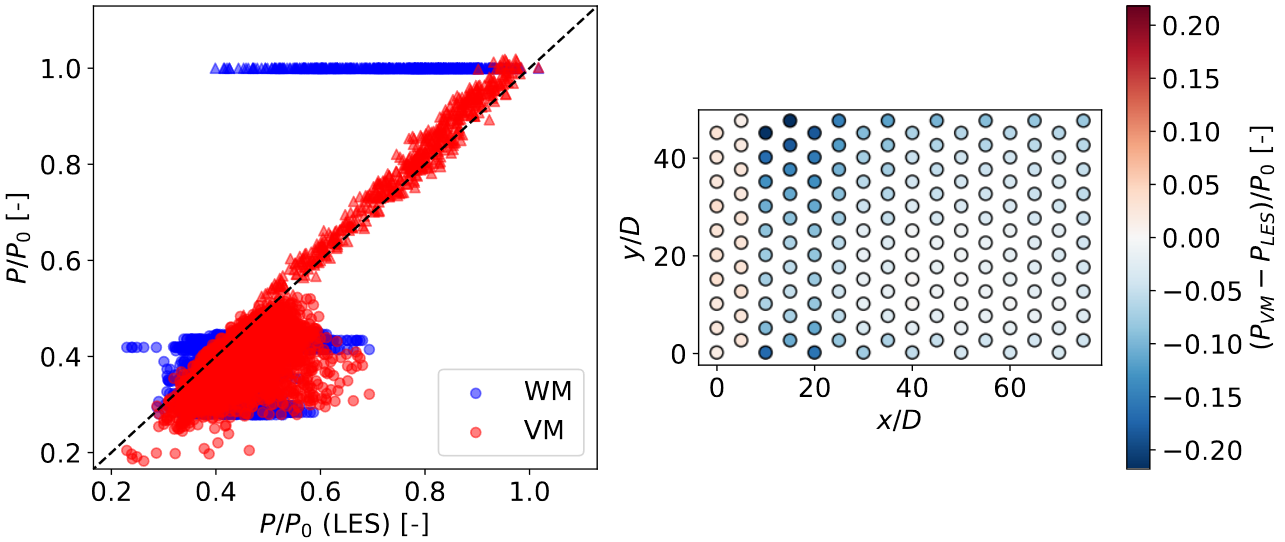}}
	\caption{Left: the turbine power outputs for all cases for both the uncoupled (blue) and coupled (red) wake model as compared to the LES. The triangles and the circles denote the first two turbine rows and the remainder of the farm, respectively. Right: the average power output error of the velocity matching method for each turbine in the farm.}
\label{fig:wmc_power}
\end{figure}

We also compare the turbine power outputs for all flow cases. The results are shown in figure \ref{fig:wmc_power}. The coupled wake model significantly outperforms the uncoupled wake model. The power output of the front row turbines is retrieved very well, with the VM method having an average error across all front turbines of 3\%, compared to 24\% for the uncoupled wake model. This shows that the coupling correctly captures the blockage effect in the mesoscale velocity fields. It also has lower errors throughout the farm when compared to the uncoupled wake model, as it takes the positive effect of the favorable pressure gradient in the farm into account. That said, it still underestimates farm power output, with the largest errors situated between the second and fifth turbine rows (see figure \ref{fig:wmc_power}, right). This is probably primarily caused by a combination of two factors. First, wake deflection, which occurs in the farm entrance region under strong stratification conditions, will cause some turbines to not be waked in the studied farm layout. As we simplify the wake model to be uni-directional in this work, we do not capture this effect. However, this is not an inherent limitation of the VM method, as it can be extended to two-directional flow using the formulation of the wake model outlined in section \ref{sec:wmc}. The right side of figure \ref{fig:wmc_power} shows that the largest errors are predicted for the turbines in the entrance region at the side of the farm, which are the turbines that are most affected by upstream wake deflection. Second, as can be seen in the left side of figure \ref{fig:nbl_wm_performance}, the wake model slightly overestimates the velocity throughout the farm. As discussed in the previous section, the VM method depends on the performance of the wake model, so more realistic wake models could improve the results. Future work should address both of these issues.

We conclude that the VM coupling method performs very well. It reproduces the mesoscale flow throughout the farm, which also results in a better approximation of the local flow. Furthermore, this translates to a better capturing of the turbine power outputs, especially for the front row turbines. This good performance was consistent across all analyzed flow cases. 

\subsection{Momentum budget analysis}
\label{sec:mba}
To investigate the importance of each of the terms in equations \ref{eq:momentum_1} and \ref{eq:momentum_2}, we perform a momentum budget analysis for the LES results of \citet{lanzilao2023parametric}. We consider the momentum balance in the streamwise direction, denoted by $x$, across the wind farm. At each position, we evaluate the momentum balance across the lateral cross-section of the wind farm, using an infinitesimally thin domain with the wind farm width $W$ and length $dx$. Note that we perform this analysis on the APM momentum equations, which also implies a horizontal filtering and height-averaging of the flow. After substracting the background momentum balance between the Coriolis force, vertical momentum flux, and geostrophic pressure gradient of the precursor simulation, the momentum balance in the lower layer becomes:
\begin{multline}
	\allowdisplaybreaks
	\underbrace{-\int_0^{W}{\overline{\boldsymbol{u}}_{h,1}\bcdot\bnabla_h\overline{u}_{1}\mathrm{d}y}}_{\mathcal{F}_{u,1}} 
	\underbrace{-\int_0^{W}{\frac{1}{\rho_0}\frac{\partial\overline{p}_1}{\partial x}\mathrm{d}y}}_{\mathcal{F}_{p,1}} 
	\underbrace{-\int_0^{W}{f_c\overline{v}_{1}^{\prime}\mathrm{d}y}}_{\mathcal{F}_{C,1}} \\ 
	+ \underbrace{\int_0^{W}{\overline{\tau}_{xx,1}\mathrm{d}y} + \left.\overline{\tau}_{xy,1}\right\vert^{W}_{0} + \int_0^{W}{\left(\frac{\overline{\tau}_{3x,1}-\overline{\tau}_{3x,0}}{h_1} - \frac{T_{3x,1}-T_{3x,0}}{H_1}\right)\mathrm{d}y}}_{\mathcal{F}_{\tau,1}} \\
	\underbrace{+\int_0^{W}{\frac{\overline{f}_{wf,x}}{h_1}\mathrm{d}y}}_{\mathcal{F}_{t}} 
	\underbrace{+\int_0^{W}{\left(\bnabla\cdot\left(\boldsymbol{\tau}_{d,x}\right)\right)_1\mathrm{d}y}		+\int_0^{W}{\frac{1}{h_1}\bnabla_h\cdot\left(h_1\left(\overline{\boldsymbol{u}}_{h,1}^{\prime\prime\prime}\overline{u}_{x,1}^{\prime\prime\prime}\right)_1\right)\mathrm{d}y}}_{\mathcal{F}_{sg,1}} \\
	+ \underbrace{\int_0^{W}{\frac{\mathcal{R}_{x,1}}{h_1}\mathrm{d}y}}_{\mathcal{R}_{1}} = 0.\label{eq:mba_1}
\end{multline}
The terms in the above equation represent the advection of momentum ($\mathcal{F}_{u}$), the pressure gradient ($\mathcal{F}_{p}$), the Coriolis force ($\mathcal{F}_{C}$), the turbulent momentum fluxes ($\mathcal{F}_{\tau}$), the turbine thrust forces ($\mathcal{F}_{t}$), the subgrid force terms ($\mathcal{F}_{sg}$), and the residual term ($\mathcal{R}$). For the upper layer, the result is analogous, but there is only a momentum flux at the bottom of the layer, and there is no wind farm thrust contribution. The above equation is the complete momentum budget, without any parametrizations or linearization.

We will now calculate these terms from the LES results of \citet{lanzilao2023parametric}. Section \ref{subsubsec:mba_les} directly calculates them as written in equation \ref{eq:mba_1}, in order to get insight into which effects are important to mesoscale wind farm flows. Afterwards, section \ref{subsubsec:mba_les} calculates them by applying the parametrizations developed in section \ref{subsec:parametrizations} to the LES-based APM states, so that these parametrizations can be validated a priori.

\subsubsection{LES data analysis}
\label{subsubsec:mba_les}
We now evaluate the momentum budget for the case H500-$\Delta\theta$5-$\Gamma$4, which is the same case as shown in section \ref{subsec:vm_performance}. It has a Froude number of approximately 1, with internal waves strong enough to prevent the choking effect described by \citet{Smith_2010}. There are weak resonant lee waves that cause velocity variations throughout the farm, and there is a moderate blockage effect.

\begin{figure}
	\centerline{\includegraphics[width=1.00\textwidth]{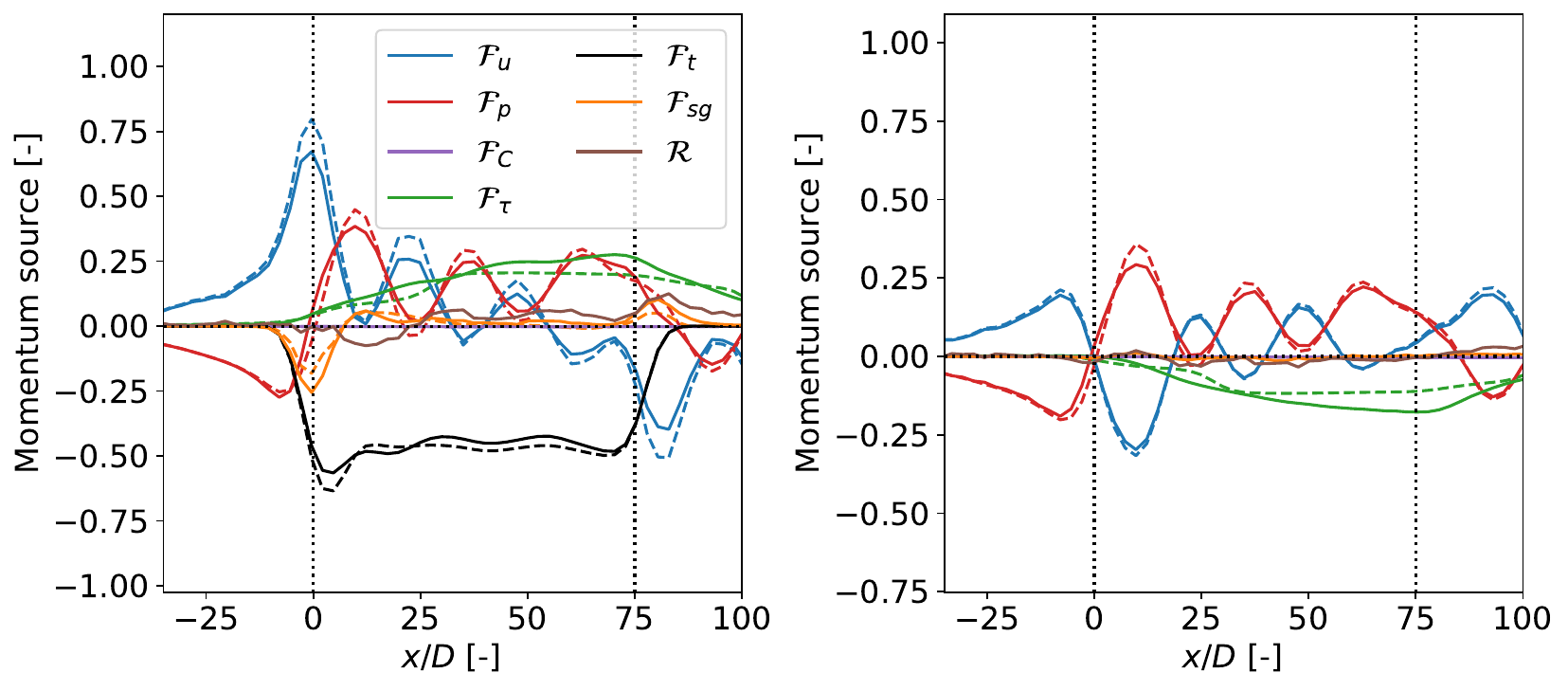}}
	\caption{Streamwise variation of all the terms in the momentum budget in the lower (left) and upper (right) layers. All the terms are scaled with $W\vert\vert\boldsymbol{U}_1\vert\vert^2/L$ and $W\vert\vert\boldsymbol{U}_2\vert\vert^2/L$ in the lower and upper layer, respectively, where $L$ is the farm length. The full lines are the terms as calculated from the LES data, and the dashed lines are the parametrized and linearized versions, calculated from the LES-based APM state. The vertical dotted lines indicate the wind farm region.}
	\label{fig:momentum_budget}
\end{figure}

Figure \ref{fig:momentum_budget} shows the streamwise momentum balance through the farm. Both the lower and upper layer are shown, and there are substantial differences in the flow dynamics.
In the lower layer, the turbine forces are felt directly, and they are balanced primarily by the convective deceleration, the pressure gradients, the subgrid forces, and the turbulent momentum fluxes. While the wind farm force is relatively constant throughout the farm, the other terms vary strongly, and which term is most important can change depending on the location. Upstream, the only active terms are the unfavourable pressure gradient and the flow deceleration. At the farm entrance, the pressure gradient becomes smaller, as the pressure perturbation reaches its maximum, and the turbine forces are completely balanced by the flow deceleration and the dispersive stresses. The rise in the dispersive velocity contributions is to be expected, as within the farm the turbine wakes form a strongly heterogeneous flow field with variations below the filter length of $L=1\mathrm{km}$. Recently, \citet{Bastankhah2023} also found that dispersive stresses played an important role in the momentum balance in wind farm flows. However, they only averaged in the lateral direction, and found the terms to contribute at turbine-length scales in the streamwise direction. In contrast, we find that when the flow is filtered in the streamwise direction as well, the main contribution of this term occurs at the start of the farm as the flow heterogeneity is rapidly established. Throughout the farm, the dispersive stresses diminish slowly as the wake mixing increases, so that their divergence at the farm exit is not as strong as their rise at the entrance. Overall, we conclude that these dispersive stresses are important to take into account when parametrizing wind farms in the APM, as their maximum effect is as strong as that of the pressure gradient.

Further downstream, the wind farm forcing is mostly counteracted by the flow acceleration, the pressure gradient, and the turbulent momentum fluxes, which primarily consist of the vertical flux contributions. Across the farm, the pressure gradient is favorable, while the turbulent momentum fluxes slowly increase. As a result, the flow gradually decelerates less throughout the farm, with the velocity reaching a roughly constant value towards the end of the farm and increasing again in the farm wake. On top of these average trends, the pressure gradient and flow convection also show strong oscillations throughout the farm, which exactly balance each other. These are the resonant lee waves described by \citet{Dries_Sensitivity_and_feedback}.

In the upper layer, the momentum balance is simpler. The main contributions are from the oscillations in the pressure gradient and the flow acceleration, as the resonant lee waves excite the whole ABL. On top of that, the turbulent momentum flux, which prevented flow deceleration and helped wake recovery in the lower layer, now slows down the flow, although it is largely counteracted by the favorable pressure gradient within the farm.

Finally, we note that the residual term $\mathcal{R}$, which is computed as the sum of all the other terms, is small compared to the terms discussed above. The only smaller term is the Coriolis contributions $\mathcal{F}_{C}$.

\subsubsection{APM approximations}
We now verify the correctness of the approximations made when linearizing and parametrizing the various terms in the APM. To this end, the dashed lines in figure \ref{fig:momentum_budget} show the momentum budget analysis of equations \ref{eq:lin_momentum_1}-\ref{eq:lin_momentum_2}. Note that this is obtained without running the APM, and instead computed using the LES-based APM state. The convective terms are linearized, the height-averaged pressure has been replaced with the pressure at the top of the ABL $p_t$, and the turbulent momentum fluxes are calculated using the linearized versions of equations \ref{eq:shear_stress_1}, \ref{eq:shear_stress_2}, and \ref{eq:tau_hor}. The wind farm forces and dispersive stresses are evaluated using the wake model, coupled to the LES-based APM state. The background state has been computed using the precursor velocity and potential temperature profiles.

Overall, the APM parametrizations perform well. The subgrid stresses are captured very accurately, and the turbulent momentum fluxes are reproduced when $\Delta\boldsymbol{\tau}_{wf}$ is included. The pressure gradient is matched very well by using $p_t$ instead of $p_1$ and $p_2$, indicating that the pressure perturbations are dominated by the gravity waves in the free atmosphere. This hydrostatic approximation is worse for the $H=1\mathrm{km}$ cases (not shown), but still holds there as well.

The main discrepancy comes from the linearization of the convective terms. As the mesoscale velocity drops, the flow deceleration is overestimated by a factor of $U_1/\overline{u}_1$. With flow perturbations of roughly $u_1^{\prime}/U_1\approx0.25$, this mismatch can be severe.

\subsection{APM validation}
\label{sec:APM_validation}
The APM is validated using the same LES data from \citet{lanzilao2023parametric} as in the previous sections. To do this, the APM was run for each of the flow cases listed in table \ref{tab:flow_cases} with the wind farm described in table \ref{tab:wind_farm}. For each flow case, the background ABL state around which the APM is linearized was based on the corresponding precursor simulation. As the APM does not have a fringe region, the periodic boundary conditions of the Fourier spectral method were dealt with by using a domain length of $L_x=\text{10 000}\mathrm{km}$. We use the same domain width of $L_y=30\mathrm{km}$, with the same periodic boundary conditions as the LES. Finally, we used a grid spacing of $\Delta x=\Delta y=500\mathrm{m}$, as was done by \citet{Dries_Sensitivity_and_feedback}.

Section \ref{subsec:flow_physics} discusses the flow fields generated by the APM, and its performance in capturing the meso-scale wind farm effects. Section \ref{subsec:power_output} investigates the power output predictions made by the APM.

\subsubsection{Flow physics}
\label{subsec:flow_physics}
To analyze the flow states produced by the APM, we take an in-depth look at the H500-$\Delta\theta$5-$\Gamma$4 case from \citet{lanzilao2023parametric}, which was also discussed in the previous sections. Both the flow physics and the APM performance are representative of the total dataset. We provide both a qualitative description of the flow phenomena, and indicate the strengths and shortcomings of the APM.

\begin{figure}
\centerline{\includegraphics[width=1.00\textwidth]{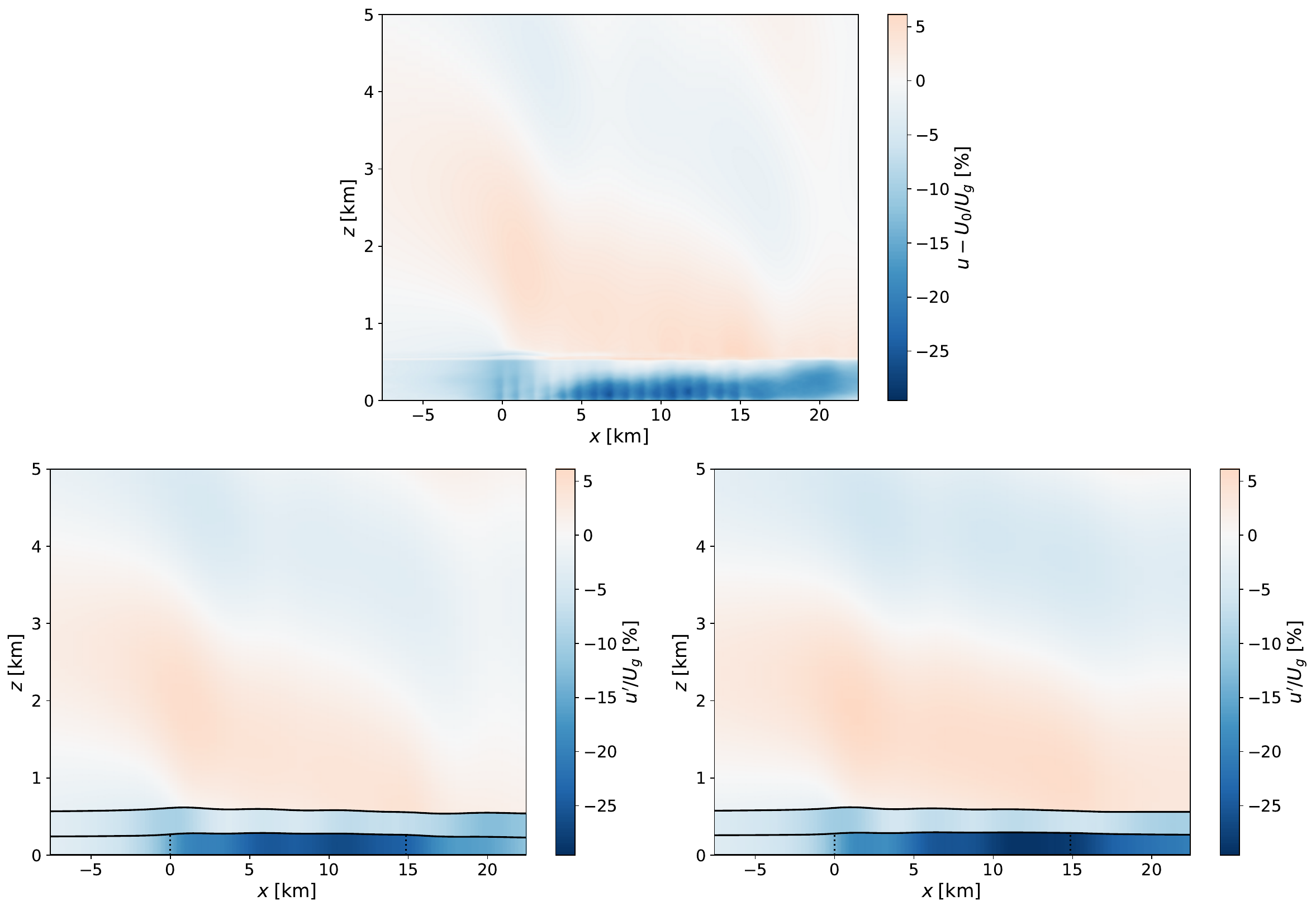}}
\caption{Top: Cross-section through the center of the farm of the velocity perturbation for the case H500-$\Delta\theta$5-$\Gamma$4, as found by LES. Bottom: Cross-section through the center of the farm of the velocity perturbation for the APM-state based on the LES results (left) and as found by the APM (right) for the case H500-$\Delta\theta$5-$\Gamma$4. The full black lines denote the pliant surfaces separating the APM layers, while the dashed lines show the wind farm region. The flow above the capping inversion is found using linear gravity wave theory.}
\label{fig:total_du_comp}
\end{figure}

To qualitatively compare the output of the APM against the LES, figure \ref{fig:total_du_comp} shows the streamwise cross-sections of the flow through the center of the farm. The capping inversion separating the ABL from the free atmosphere is clearly visible. The displacement of this capping inversion triggers internal gravity waves in the free atmosphere above it, and these along with the pressure feedback from the inversion displacement itself cause a significant velocity reduction upstream of the farm. The gravity waves within the free atmosphere are captured fairly well up to 5km. Within the ABL, the resonant lee waves described by \citet{Dries_Sensitivity_and_feedback} are visible, especially above the internal boundary layer (IBL) forming above the wind farm. Notably, the APM states do not capture this IBL, as it only contains the height-averaged flow within the layers. While the APM models the flow upstream and throughout the farm fairly well, it does not accurately represent the farm wake structure. This is to be expected, due to its inability to model IBL growth and limited turbulence parametrization. Despite these shortcomings, the comparison shows decent agreement between the APM and LES, especially for the stratification-related flow features.

\begin{figure}
	\centerline{\includegraphics[width=1.00\textwidth]{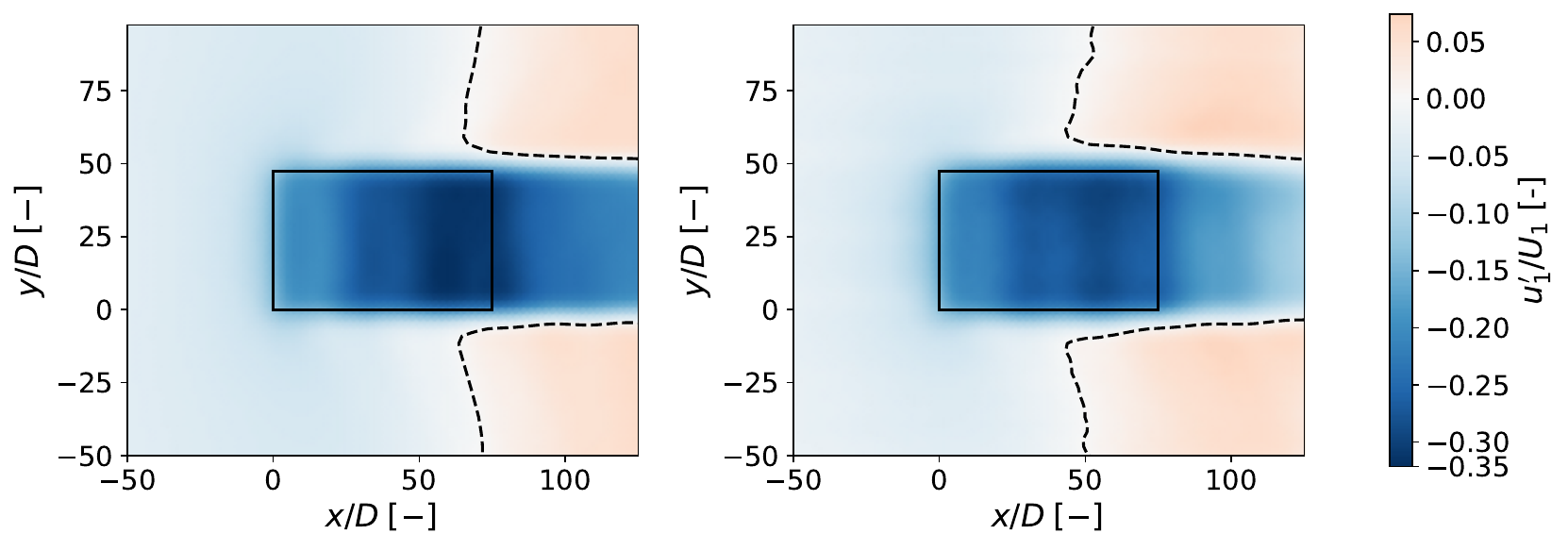}}
	\caption{Top view of the lower layer velocity perturbation in the $x$-direction for the APM (left) and the LES (right) output for the case H500-$\Delta\theta$5-$\Gamma$4. The full black rectangles show the wind farm region. The dashed black lines show the zero contour.}
	\label{fig:u_field_comp}
\end{figure}

Continuing this qualitative discussion, figure \ref{fig:u_field_comp} shows the top-view of the $u_1^{\prime}$ perturbation velocity for both the APM and the LES-based state. Once again, the blockage effect in front of the farm is clearly seen. From this top-view, the resonant lee waves are again visible. As noted by \citet{lanzilao2023parametric}, linear theory accurately predicts the length scale of these waves, as well as the angles of the characteristic lines emanating from the farm (not visible). While the agreement for the lee waves' length scales is good, the APM does not consistently reproduce their location in all other flow cases, most notably for the H300 cases with strong capping inversions and weak upper atmosphere stratification (a complete comparison of the lower-layer mesoscale velocity profiles through the center of the farm can be found in appendix \ref{app:additional_plots}). Additionally, figure \ref{fig:u_field_comp} also shows the speed-up of the flow around the farm, which is caused by the pressure gradient of the gravity waves. However, the location of where the velocity perturbation becomes positive, shown in dashed lines, is located too far downstream when compared to the LES.

\begin{figure}
	\centerline{\includegraphics[width=1.00\textwidth]{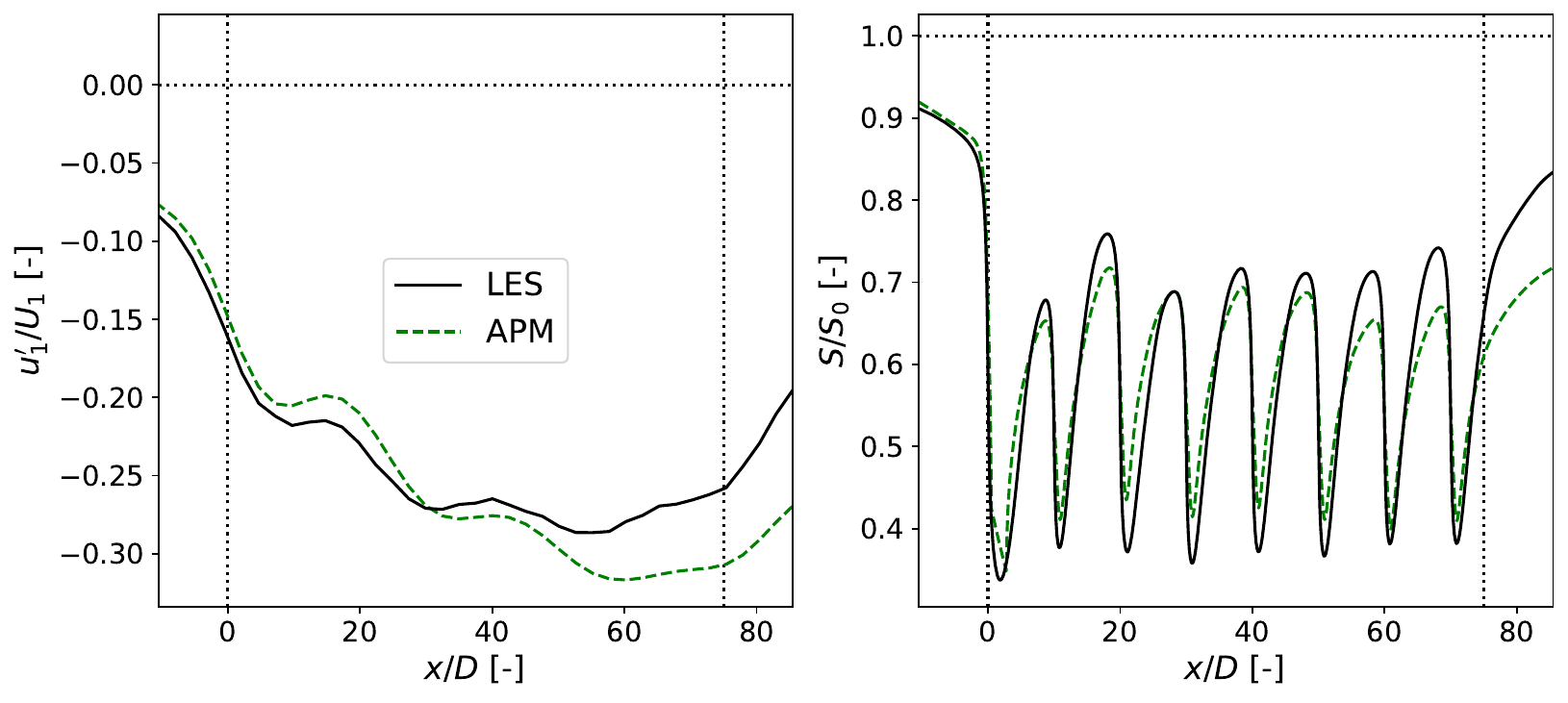}}
	\caption{Streamwise velocity deficits through the center of the farm of the LES (full, black) and APM output (dashed, green) for the case H500-$\Delta\theta$5-$\Gamma$4. Left: the meso-scale velocity deficit. Right: the average velocity across a tube with the turbine diameter. The dotted lines denote the wind farm region.}
	\label{fig:tlm_velocity}
\end{figure}

Finally, by comparing the velocity deficits in figure \ref{fig:u_field_comp}, one sees that the APM slightly underpredicts the velocity deficit in the first half of the farm, and strongly overpredicts it in the second half and the farm wake. To see this more accurately, the left side of figure \ref{fig:tlm_velocity} shows the difference between the centerline velocity deficit for both the LES and APM results. At the farm entrance, the APM captures the blockage fairly well, although the predicted velocity is slightly high. Halfway through the farm, the APM predicts that the velocity deficit increases drastically, resulting in a lower velocity than the LES. As a result, the APM overpredicts the farm wake as discussed above. The discrepancies in the velocity estimate are reflected in the turbine-level velocity, as shown in the right side of figure \ref{fig:tlm_velocity}. This is consistent across most analyzed cases, as shown in Appendix \ref{app:additional_plots}, except for some H300 cases where the location of strong resonant lee waves doesn't match the LES.

Overall, we conclude that the APM captures the relevant physics well, and produces realistic flow fields. It is worth noting that this good qualitative agreement between the APM and LES has been achieved with minimal tuning. The only parameters fitted in this work are the ones describing the wind-farm induced turbulent momentum flux.

\subsubsection{Power output}
\label{subsec:power_output}
\begin{figure}
	\centerline{\includegraphics[width=1.00\textwidth]{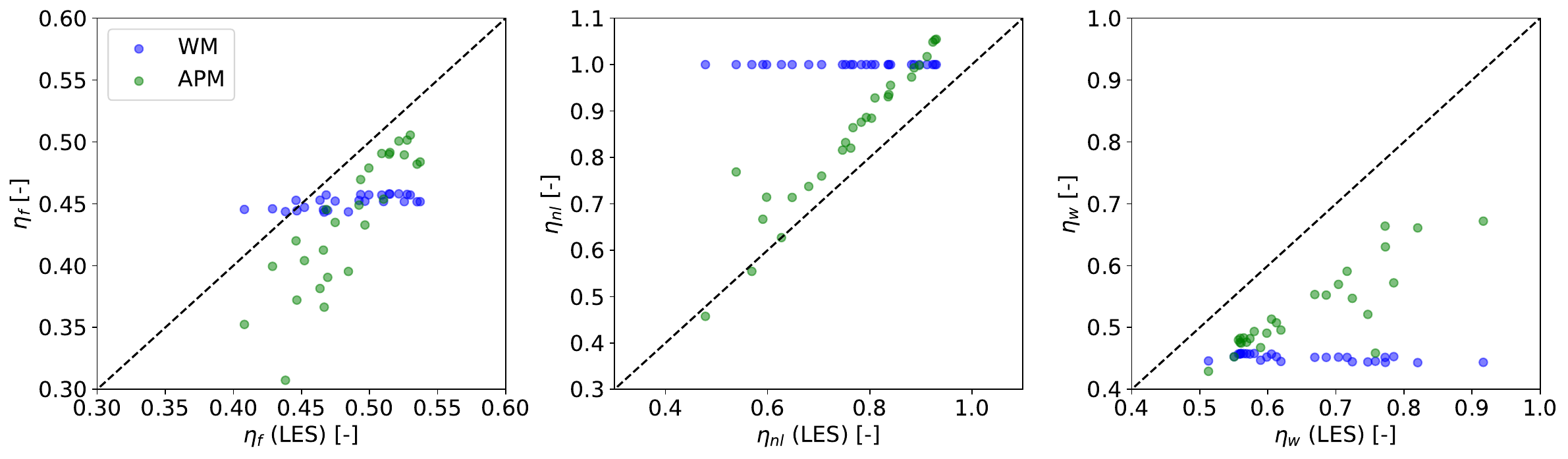}}
	\caption{Comparison of the farm (left), non-local (middle), and wake (right) efficiencies as found by LES (horizontal axes) for all cases with those predicted by the APM (green) and the uncoupled wake model (blue).}
	\label{fig:efficiencies}
\end{figure}

The aim of the APM is to provide the power predictions that include the effects of blockage and other meso-scale phenomena on wind farm performance. These are farm-wide effects, and are therefore best quantified using farm-wide metrics. We follow \citet{Dries_Gravity_waves} by using the non-local wind-farm efficiency $\eta_{nl}$, which is the ratio of the average power output of the first row turbines and an identical turbine operating on its own:
\begin{equation}
	\eta_{nl} = \frac{P_1}{P_0}.
	\label{eq:eta_nl}
\end{equation}
This $P_0$ is the same power output that was used to scale the results in section \ref{subsec:vm_performance}. The non-local efficiency captures the decrease in power output caused by the upstream effects of the turbines operating in a farm. Within the studied data-set, there are large variations in $\eta_{nl}$, with values going from roughly 1 for cases with minimal blockage to as low as 0.55 for some cases. Additionally, following \citet{Dries_Gravity_waves}, we define:
\begin{equation}
	\eta_f=\eta_{nl}\eta_w, \quad \eta_w=\frac{P_{avg}}{P_1},
	\label{eq:eta_fw}
\end{equation}
where $\eta_f$ is the total farm efficiency, and $\eta_w$ is the wake efficiency, defined as the ratio between the average power output of all turbines $P_{avg}$ to that of the front row average $P_1$. The wake efficiency is a measure for the turbine interactions within the wind farm, while the farm efficiency shows the performance of a wind farm as a whole. Figure \ref{fig:efficiencies} shows the farm, non-local, and wake efficiencies of the studied wind farm for all cases, as obtained by LES, the uncoupled wake model, and the APM.

The uncoupled wake model performs quite poorly, and its outputs do not vary significantly between the different flow cases. This is to be expected, as the main difference between the flow cases are the stratification levels in the capping inversion an the free atmosphere, which the wake model does not take into account. As a result, the wake model can not capture the large variations and trends in the different efficiencies. The slight variance in its predictions is due to small differences in the wind veer and shear, and turbulence intensity at hub height. Additionally, it does not model any upstream effects, so $\eta_{nl}$ is always 1. This is offset by a consistent underestimation of the wake efficiency. While this results in a roughly correct average farm efficiency when averaging over all cases, it's clear that this is because the errors in $\eta_{nl}$ and $\eta_w$ cancel each other out.

In contrast, the APM captures the trends of both $\eta_f$ and $\eta_{nl}$ quite well. This is expected after section \ref{subsec:flow_physics}, which found that the APM captures the general behavior of the atmospheric flow, and section \ref{subsec:vm_performance}, which showed that the VM method can provide good estimates for the power output. Nevertheless, as discussed in section \ref{subsec:flow_physics}, the APM consistently slightly overestimates the velocity in the entrance region. This results in an overestimation of the non-local efficiency as well, leading to an offset between the APM and LES, and an average error of 9\%. This is better than the uncoupled wake model, which on average overestimates $\eta_{nl}$ by 24\%. For four of the H1000 cases, where blockage effects are very weak, the APM predicts a non-local efficiency above one, as its velocity deficit becomes smaller than that of the uncoupled wake model. In contrast, the wake efficiency is consistently underestimated, although the APM still significantly outperforms the uncoupled wake model. The discrepancies in $\eta_w$ are mainly due to the overestimation of $P_1$, the underestimation of the velocity in the second half of the farm, and the issues with the VM method found in the previous section.

\begin{figure}
	\centerline{\includegraphics[width=0.6\textwidth]{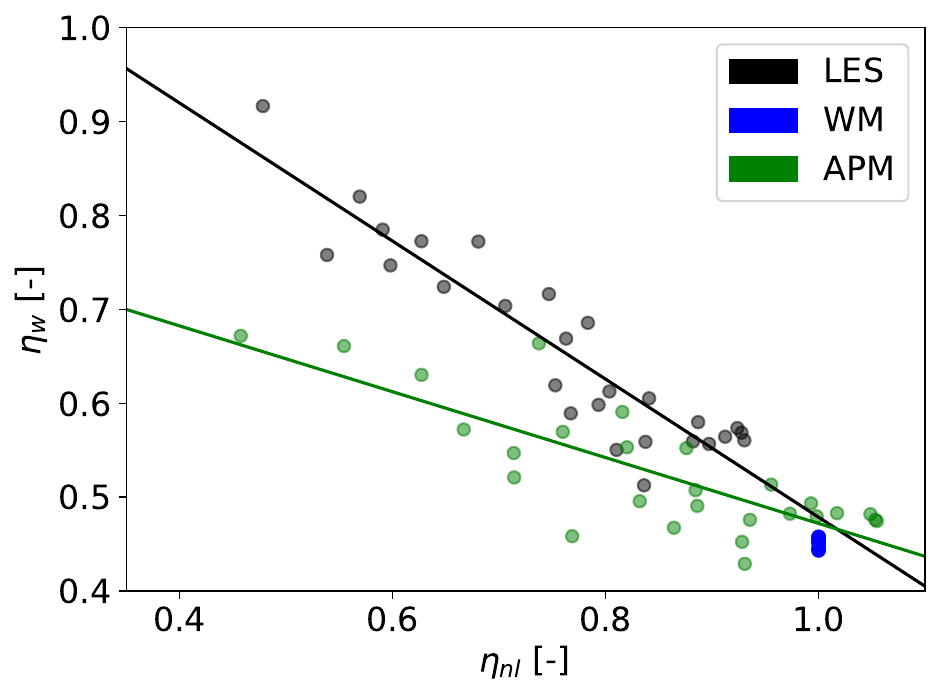}}
	\caption{Relation between $\eta_{nl}$ and $\eta_w$, as found by LES (black), the APM (green), and the uncoupled wake model (blue). The lines represent least-squares fits.}
	\label{fig:nl_effect}
\end{figure}

The relation between the unfavorable pressure gradient upstream of the farm and the favorable one throughout it, as reported by \citet{lanzilao2023parametric}, results in a relation between $\eta_{nl}$ and $\eta_w$ as well. Specifically, as $\eta_{nl}$ decreases due to higher blockage, $\eta_w$ increases, as shown in figure \ref{fig:nl_effect}. Despite this balancing effect, blockage is still detrimental to the overall power output of the farm. Figure \ref{fig:nl_effect} also shows that the APM can reproduce these trends, although the slopes are not exactly the same.

In conclusion, the APM is able to predict the variation of wind farm performance across different flow cases. More importantly, it consistently and significantly outperforms the uncoupled wake model, which is the default approach taken by the wind energy community. Finally, we again want to emphasize that this has been achieved without any power-based tuning of the model parameters.

\section{Conclusion}
\label{sec:conclusions}
This work presents a new version of the atmospheric perturbation model first introduced by \citet{Dries_Sensitivity_and_feedback}, and improves the mesoscale parametrization of large wind farms. This was mainly done in three ways. First, we added an explicit ad-hoc parametrization of the increased momentum entrainment above a wind farm, which allows for general turbine layouts. Second, we presented a large-scale overhaul of the wake model coupling. For this, a new method was developed based on ensuring that the velocity fields of the APM and the wake model were consistent with each other. Third, we found that dispersive stresses play an important role at the farm entrance, and significantly contribute to the global blockage effect. Using the wake model coupling, these stresses can be easily incorporated into the APM.

The velocity matching method was validated independently of the APM using a dataset of 27 LES simulations set up by \citet{lanzilao2023parametric}. The matching of the meso-scale velocity fields resulted in a better approximation of the turbine-level velocity fields, and drastically improved the power predictions when compared to the uncoupled wake model. This good performance was consistent, with the coupled wake model outperforming its uncoupled counterpart for all analyzed flow cases. Additionally, we identified the main sources of error as being the uni-directional approximation, and the accuracy of the wake models in representing the turbine wakes. The former could easily be addressed in future work, as the VM method is straightforward to extend to two-directional flow.

Additionally, we performed a momentum budget analysis on the LES results, and found that the dispersive stresses are important to mesoscale wind farm flows. At the farm entrance, their effect can be as strong as the peak pressure gradient, significantly contributing to the global blockage effect.  Since the APM filter results in a similar resolution to numerical weather models, dispersive stresses are presumably also important to incorporate in wind-farm parametrizations for those models, and we recommend this for future research.

Finally, the improved APM was validated using the same LES data. The APM performs well, as it captures most of the relevant meso-scale flow phenomena, especially when it comes to stratification effects. There is good qualitative agreement between the LES and the APM for the appearance of gravity waves, and the associated blockage, farm pressure gradients, resonant lee waves, and flow speed-up around the sides of the farm. However, the reduced-order nature of the APM prevents it from capturing the IBL growth, and changes in the ABL vertical structure. Moreover, the APM slightly overestimates the velocity in the first half of the farm, and underestimates it in the second half. Despite this mismatch, the good qualitative match allows the APM to model the effect of atmospheric stratification on turbine power output. For all analyzed metrics, the APM significantly outperforms a standard engineering wake model. Regarding the effect of blockage on the front row turbines, the APM overestimates the wind-farm's non-local efficiency, but reproduces the variation in this efficiency across all cases very well. This allows it to find the same relation between the non-local and the wake and farm efficiencies as the LES. Given the good performance of the underlying VM method, we expect any further improvement to the APM to translate directly into more accurate power predictions. Additionally, this was achieved without any power-based tuning. This makes the APM a promising tool for studying the next generation of off-shore wind farms.

Currently, the APM can still only simulate wind farms in highly idealized conditions. In reality, the simple CNBL profiles used in this study are complicated by various phenomena, such as ABL stability effects or meso-scale systems already present in the atmosphere. Moreover, both atmospheric conditions and wind farm operational settings are rarely steady-state, with transient effects being important for control problems. Finally, the current turbulence parametrization does not include the effects of background turbulence intensity, and could be improved by an explicit modeling of turbulent transport phenomena. Future work should extend the APM to include these flow features. The model should also be further compared to experimental and operational data.

\backsection[Supplementary data]{\label{SupMat}Not applicable.}

\backsection[Acknowledgements]{The computational resources and services in this work were provided by the VSC (Flemish Supercomputer Center), funded by the Research Foundation Flanders (FWO) and the Flemish Government department EWI.}

\backsection[Funding]{This research has been supported by the Energy Transition Fund of the Belgian Federal Public Service for Economy, SMEs, and Energy (FOD Economie, K.M.O., Middenstand en Energie), and by the European Union Horizon Europe Framework programme (HORIZON-CL5-2021-D3-03-04) under grant agreement no. 101084205.}

\backsection[Declaration of interests]{The authors report no conflict of interest.}

\backsection[Data availability statement]{An open-source version of the code will be released with the final version of this paper. An earlier version of the model can be found at https://doi.org/10.48804/XMNVVY.}

\backsection[Author ORCIDs]{K. Devesse, https://orcid.org/0000-0003-2404-6444; L. Lanzilao, https://orcid.org/0000-0003-1976-3449; J. Meyers, https://orcid.org/0000-0002-2828-4397}

\backsection[Author contributions]{KD and JM jointly rederived the model, and incorporated the additional linear terms, dispersive stresses, and the increased momentum flux into the APM. KD, LL, and JM jointly developed the velocity matching coupling method. KD and JM jointly set up the validation studies. KD and LL performed code implementations and carried out the simulations. KD, LL, and JM jointly wrote the manuscript.}

\appendix

\section{Wind-farm induced momentum flux tuning}    
\label{app:tau_wf_fit}
\begin{figure}
	\centerline{\includegraphics[width=0.6\textwidth]{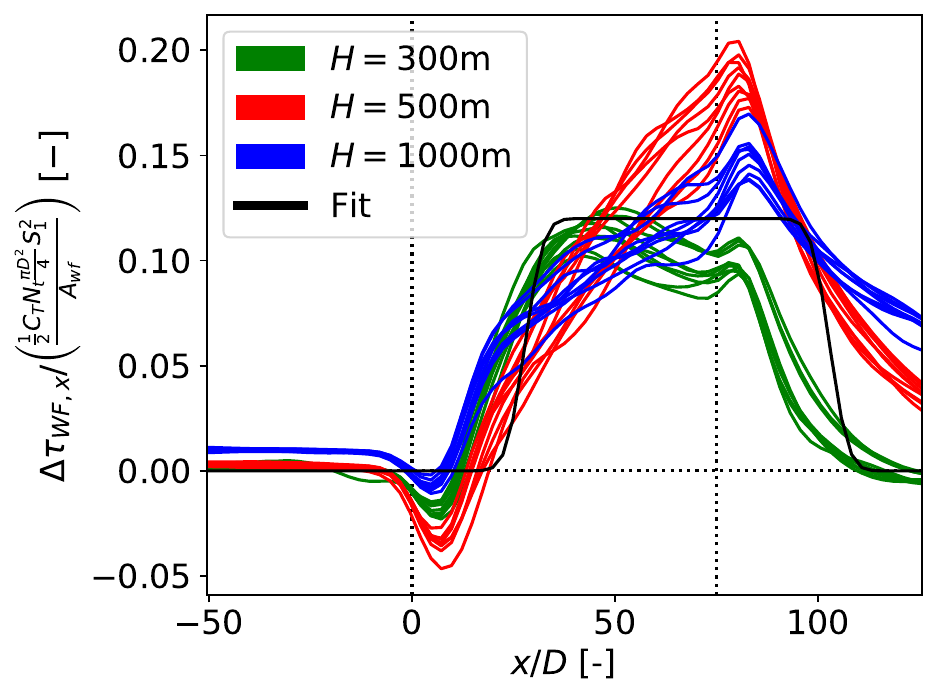}}
	\caption{Added wind farm momentum flux $\Delta\overline{\boldsymbol{\tau}}_{\textit{wf}}$ as calculated based on the LES data of \citet{lanzilao2023parametric}. The line colors denote the boundary layer height. The fitted profile is indicated by the black line. The dotted lines indicate the wind farm region.}
	\label{fig:dtau_wf}
\end{figure}

We fit the coefficients $a_{\tau}$ and $d_{\tau}$ to the 27 LES cases from \citet{lanzilao2023parametric} used throughout this paper and summarized in tables \ref{tab:flow_cases} and \ref{tab:wind_farm}. From this dataset, we have constructed LES-based APM states, as described in section \ref{sec:validation}. This way, $\overline{\boldsymbol{u}}_1$ and $\overline{\boldsymbol{u}}_2$ are known for all cases. Based on the associated precursor simulations, we compute $D$. We then apply the Gaussian filter described in equation \ref{eq:gaussian_kernel} to the vertical momentum fluxes, and evaluate $\overline{\tau}_{03}$ at $z_1$ for all cases. This gives us data for the first two terms in equation \ref{eq:shear_stress_2}, allowing us to calculate $\Delta\boldsymbol{\tau}_{\textit{wf}}$ directly from the LES. We average the resulting fields along the spanwise direction within the wind farm boundaries, and fit the coefficients along the streamwise direction. Figure \ref{fig:dtau_wf} shows the LES profiles and the resulting fit for $\Delta\overline{\boldsymbol{\tau}}_{\textit{wf}}$, Gaussian-filtered onto the APM grid.

\section{Velocity comparisons for all analyzed cases}    
\label{app:additional_plots}
\begin{figure}
	\centerline{\includegraphics[width=0.8\textwidth]{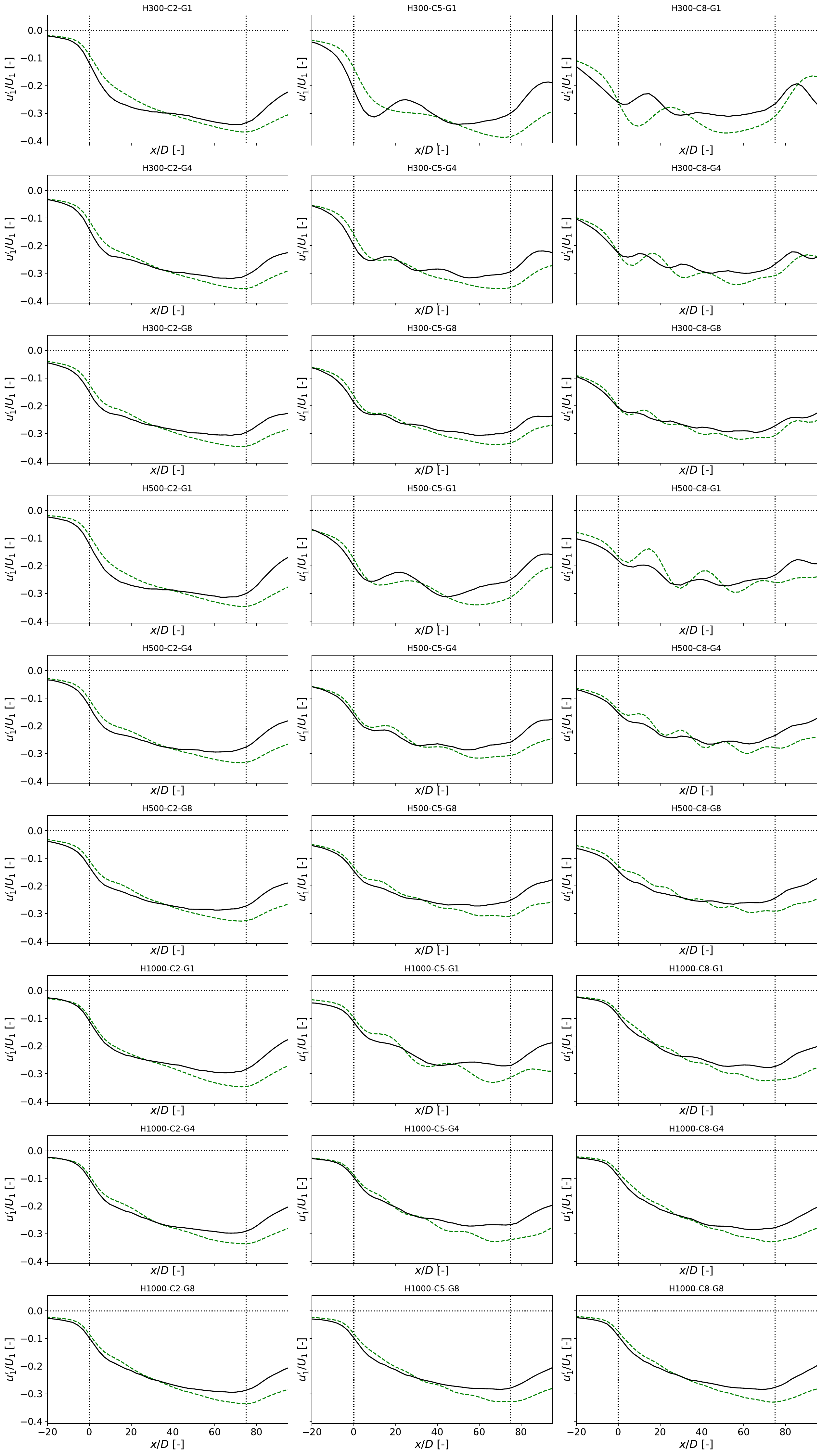}}
	\caption{Lower layer velocity perturbations through the center of the farm of the LES (full, black) and APM output (dashed, green) for all analyzed cases. The dotted lines denote the wind farm region.}
	\label{fig:LES_APM_comparison}
\end{figure}

Figure \ref{fig:LES_APM_comparison} shows the velocity deficits along the centerline of the farm for all analyzed flow cases. The APM consistently over- and underpredicts the velocity in the first and second half of the farm respectively, but otherwise captures the stratification effects quite well. The largest errors occur for the cases with low, strong capping inversions and weak atmospheric stratification above.

\bibliographystyle{jfm}

\begin{thebibliography}{51}
	\expandafter\ifx\csname natexlab\endcsname\relax\def\natexlab#1{#1}\fi
	\def\au#1{#1} \def\ed#1{#1} \def\yr#1{#1}\def\at#1{#1}\def\jt#1{\textit{#1}}
	\def\bt#1{#1}\def\bvol#1{\textbf{#1}} \def\vol#1{#1} \def\pg#1{#1}
	\def\publ#1{#1}\def\arxiv#1{#1}\def\org#1{#1}\def\st#1{\textit{#1}}
	
	\bibitem[Abkar \& Port{\'{e}}-Agel(2013)]{Abkar2013}
	{\sc \au{Abkar, Mahdi} \& \au{Port{\'{e}}-Agel, Fernando}} \yr{2013}  \at{{The
			Effect of Free-Atmosphere Stratification on Boundary-Layer Flow and Power
			Output from Very Large Wind Farms}}.  \jt{Energies}  \bvol{6}~(5),
	\pg{2338--2361}.
	
	\bibitem[Allaerts {\em et~al.\/}(2018)Allaerts, Broucke, {Van Lipzig} \&
	Meyers]{Dries_Annual_Impact}
	{\sc \au{Allaerts, Dries}, \au{Broucke, Sam~Vanden}, \au{{Van Lipzig}, Nicole}
		\& \au{Meyers, Johan}} \yr{2018}  \at{{Annual impact of wind-farm gravity
			waves on the Belgian-Dutch offshore wind-farm cluster}}.  \jt{Journal of
		Physics: Conference Series}  \bvol{1037}~(7).
	
	\bibitem[Allaerts \& Meyers(2017)]{Dries_Boundary_layer_development}
	{\sc \au{Allaerts, Dries} \& \au{Meyers, Johan}} \yr{2017}  \at{{Boundary-layer
			development and gravity waves in conventionally neutral wind farms}}.
	\jt{Journal of Fluid Mechanics}  \bvol{814},  \pg{95--130}.
	
	\bibitem[Allaerts \& Meyers(2018)]{Dries_Gravity_waves}
	{\sc \au{Allaerts, Dries} \& \au{Meyers, Johan}} \yr{2018}  \at{{Gravity Waves
			and Wind-Farm Efficiency in Neutral and Stable Conditions}}.
	\jt{Boundary-Layer Meteorology}  \bvol{166}~(2),  \pg{269--299}.
	
	\bibitem[Allaerts \& Meyers(2019)]{Dries_Sensitivity_and_feedback}
	{\sc \au{Allaerts, Dries} \& \au{Meyers, Johan}} \yr{2019}  \at{{Sensitivity
			and feedback of wind-farm-induced gravity waves}}.  \jt{Journal of Fluid
		Mechanics}  \bvol{862},  \pg{990--1028}.
	
	\bibitem[Bastankhah {\em et~al.\/}(2023)Bastankhah, Mohammadi, Lees, Diaz,
	Buxton \& Ivanell]{Bastankhah2023}
	{\sc \au{Bastankhah, Majid}, \au{Mohammadi, Mohammad~Mehdi}, \au{Lees,
			Charlie}, \au{Diaz, Gonzalo Pablo~Navarro}, \au{Buxton, Oliver} \&
		\au{Ivanell, Stefan}} \yr{2023}  \at{{A fast-running physics-based wake model
			for a semi-infinite wind farm}}  \pg{pp. 1--32},  \arxiv{arXiv: 2309.08711}.
	
	\bibitem[Bastankhah \& Port{\'{e}}-Agel(2014)]{bastankhah2014new}
	{\sc \au{Bastankhah, Majid} \& \au{Port{\'{e}}-Agel, Fernando}} \yr{2014}
	\at{{A new analytical model for wind-turbine wakes}}.  \jt{Renewable Energy}
	\bvol{70},  \pg{116--123}.
	
	\bibitem[Bleeg {\em et~al.\/}(2018)Bleeg, Purcell, Ruisi \&
	Traiger]{bleeg2018wind}
	{\sc \au{Bleeg, James}, \au{Purcell, Mark}, \au{Ruisi, Renzo} \& \au{Traiger,
			Elizabeth}} \yr{2018}  \at{{Wind farm blockage and the consequences of
			neglecting its impact on energy production}}.  \jt{Energies}  \bvol{11}~(6).
	
	\bibitem[Bortolotti {\em et~al.\/}(2019)Bortolotti, Tarres, Dykes, Merz,
	Sethuraman, Verelst \& Zahle]{bortolotti2019iea}
	{\sc \au{Bortolotti, Pietro}, \au{Tarres, Helena~C}, \au{Dykes, Katherine~L},
		\au{Merz, Karl}, \au{Sethuraman, Latha}, \au{Verelst, David} \& \au{Zahle,
			Frederik}} \yr{2019}  \at{Iea wind tcp task 37: Systems engineering in wind
		energy - wp2.1 reference wind turbines} .
	
	\bibitem[Branlard {\em et~al.\/}(2020)Branlard, Quon, {Meyer Forsting}, King \&
	Moriarty]{Branlard2020}
	{\sc \au{Branlard, Emmanuel}, \au{Quon, Eliot}, \au{{Meyer Forsting},
			Alexander~R.}, \au{King, Jennifer} \& \au{Moriarty, Patrick}} \yr{2020}
	\at{{Wind farm blockage effects: comparison of different engineering
			models}}.  \jt{Journal of Physics: Conference Series}  \bvol{1618}~(6),
	\pg{062036}.
	
	\bibitem[Calaf {\em et~al.\/}(2010)Calaf, Meneveau \& Meyers]{Calaf2010}
	{\sc \au{Calaf, Marc}, \au{Meneveau, Charles} \& \au{Meyers, Johan}} \yr{2010}
	\at{{Large eddy simulation study of fully developed wind-turbine array
			boundary layers}}.  \jt{Physics of Fluids}  \bvol{22}~(1),  \pg{015110}.
	
	\bibitem[Centurelli {\em et~al.\/}(2021)Centurelli, Vollmer, Schmidt,
	D{\"{o}}renk{\"{a}}mper, Schr{\"{o}}der, Lukassen \& Peinke]{Centurelli2021}
	{\sc \au{Centurelli, G.}, \au{Vollmer, L.}, \au{Schmidt, J.},
		\au{D{\"{o}}renk{\"{a}}mper, M}, \au{Schr{\"{o}}der, M.}, \au{Lukassen,
			L.~J.} \& \au{Peinke, J.}} \yr{2021}  \at{{Evaluating Global Blockage
			engineering parametrizations with LES}}.  \jt{Journal of Physics: Conference
		Series}  \bvol{1934}~(1),  \pg{012021},  \arxiv{arXiv: 2103.10908}.
	
	\bibitem[Csanady(1974)]{Csanady1974}
	{\sc \au{Csanady, G.~T.}} \yr{1974}  \at{{Equilibrium theory of the planetary
			boundary layer with an inversion lid}}.  \jt{Boundary-Layer Meteorology}
	\bvol{6}~(1-2),  \pg{63--79}.
	
	\bibitem[Devesse {\em et~al.\/}(2022)Devesse, Lanzilao, Jamaer, {Van Lipzig} \&
	Meyers]{Devesse2022}
	{\sc \au{Devesse, Koen}, \au{Lanzilao, Luca}, \au{Jamaer, Sebastiaan}, \au{{Van
				Lipzig}, Nicole} \& \au{Meyers, Johan}} \yr{2022}  \at{{Including realistic
			upper atmospheres in a wind-farm gravity-wave model}}.  \jt{Wind Energy
		Science}  \bvol{7}~(4),  \pg{1367--1382}.
	
	\bibitem[Doekemeijer {\em et~al.\/}(2022)Doekemeijer, Simley \&
	Fleming]{Doekemeijer2022}
	{\sc \au{Doekemeijer, Bart~Matthijs}, \au{Simley, Eric} \& \au{Fleming, Paul}}
	\yr{2022}  \at{{Comparison of the Gaussian Wind Farm Model with Historical
			Data of Three Offshore Wind Farms}}.  \jt{Energies}  \bvol{15}~(6),
	\pg{1964}.
	
	\bibitem[Durran(1990)]{Durran_1990}
	{\sc \au{Durran, Dale~R.}} \yr{1990} {\em Mountain Waves and Downslope
		Winds\/},  \pg{pp. 59--81}.  \publ{Boston, MA: American Meteorological
		Society}.
	
	\bibitem[Emeis(2009)]{Emeis2009}
	{\sc \au{Emeis, S.}} \yr{2009}  \at{{A simple analytical wind park model
			considering atmospheric stability}}.  \jt{Wind Energy}  \bvol{13}~(5),
	\pg{459--469}.
	
	\bibitem[Fischereit {\em et~al.\/}(2021)Fischereit, Brown, Lars{\'{e}}n, Badger
	\& Hawkes]{Fischereit2021}
	{\sc \au{Fischereit, Jana}, \au{Brown, Roy}, \au{Lars{\'{e}}n, Xiaoli~Guo},
		\au{Badger, Jake} \& \au{Hawkes, Graham}} \yr{2021}  \at{{Review of Mesoscale
			Wind-Farm Parametrizations and Their Applications}}.  \jt{Boundary-Layer
		Meteorology} .
	
	\bibitem[Frandsen(1992)]{Frandsen1992}
	{\sc \au{Frandsen, Sten}} \yr{1992}  \at{{On the wind speed reduction in the
			center of large clusters of wind turbines}}.  \jt{Journal of Wind Engineering
		and Industrial Aerodynamics}  \bvol{39}~(1-3),  \pg{251--265}.
	
	\bibitem[Frandsen {\em et~al.\/}(2006)Frandsen, Barthelmie, Pryor, Rathmann,
	Larsen, H{\o}jstrup \& Th{\o}gersen]{Frandsen2006}
	{\sc \au{Frandsen, Sten}, \au{Barthelmie, Rebecca}, \au{Pryor, Sara},
		\au{Rathmann, Ole}, \au{Larsen, S{\o}ren}, \au{H{\o}jstrup, J{\o}rgen} \&
		\au{Th{\o}gersen, Morten}} \yr{2006}  \at{{Analytical modelling of wind speed
			deficit in large offshore wind farms}}.  \jt{Wind Energy}  \bvol{9}~(1-2),
	\pg{39--53}.
	
	\bibitem[Gill(1982)]{Adrian_Gill_Atmosphere_Ocean_dynamics}
	{\sc \au{Gill, Adrian~E}} \yr{1982} {\em Atmosphere-Ocean Dynamics\/}.
	\publ{San Diego, USA: Academic Press}.
	
	\bibitem[Kirby {\em et~al.\/}(2023)Kirby, Dunstan \& Nishino]{Kirby2023}
	{\sc \au{Kirby, Andrew}, \au{Dunstan, Thomas~D} \& \au{Nishino, Takafumi}}
	\yr{2023}  \at{{An analytical model of momentum availability for predicting
			large wind farm power}}  \pg{pp. 1--19},  \arxiv{arXiv: 2306.08088}.
	
	\bibitem[Kirby {\em et~al.\/}(2022)Kirby, Nishino \& Dunstan]{Kirby2022}
	{\sc \au{Kirby, Andrew}, \au{Nishino, Takafumi} \& \au{Dunstan, Thomas~D.}}
	\yr{2022}  \at{{Two-scale interaction of wake and blockage effects in large
			wind farms}}.  \jt{Journal of Fluid Mechanics}  \bvol{953},  \pg{1--28},
	\arxiv{arXiv: 2207.03148}.
	
	\bibitem[Klemp \& Lilly(1975)]{klemp1975dynamics}
	{\sc \au{Klemp, Joseph~B.} \& \au{Lilly, D.~K.}} \yr{1975}  \at{{Dynamics of
			Wave-Induced Downslope Winds.}}  \jt{Journal of the Atmospheric Sciences}
	\bvol{32}~(2),  \pg{320--339}.
	
	\bibitem[Lam {\em et~al.\/}(2015)Lam, Pitrou \& Seibert]{numba}
	{\sc \au{Lam, Siu~Kwan}, \au{Pitrou, Antoine} \& \au{Seibert, Stanley}}
	\yr{2015} Numba: A llvm-based python jit compiler.  \bt{In {\em Proceedings
			of the Second Workshop on the LLVM Compiler Infrastructure in HPC\/}}.
	\publ{New York, NY, USA: Association for Computing Machinery}.
	
	\bibitem[Lanzilao \& Meyers(2021{\natexlab{{\em a\/}}})]{Lanzilao2021merging}
	{\sc \au{Lanzilao, Luca} \& \au{Meyers, Johan}} \yr{2021{\natexlab{{\em a\/}}}}
	\at{{A new wake-merging method for wind-farm power prediction in the
			presence of heterogeneous background velocity fields}}.  \jt{Wind Energy} ,
	\arxiv{arXiv: 2010.03873}.
	
	\bibitem[Lanzilao \& Meyers(2021{\natexlab{{\em b\/}}})]{Lanzilao2021setpoint}
	{\sc \au{Lanzilao, Luca} \& \au{Meyers, Johan}} \yr{2021{\natexlab{{\em b\/}}}}
	\at{{Set-point optimization in wind farms to mitigate effects of flow
			blockage induced by atmospheric gravity waves}}.  \jt{Wind Energy Science}
	\bvol{6},  \pg{247--271}.
	
	\bibitem[Lanzilao \& Meyers(2022)]{Lanzilao2022}
	{\sc \au{Lanzilao, Luca} \& \au{Meyers, Johan}} \yr{2022}  \at{{Effects of
			self-induced gravity waves on finite wind-farm operations using a large-eddy
			simulation framework}}.  \jt{Journal of Physics: Conference Series}
	\bvol{2265}~(2).
	
	\bibitem[Lanzilao \& Meyers(2023{\natexlab{{\em a\/}}})]{Lanzilao2023}
	{\sc \au{Lanzilao, Luca} \& \au{Meyers, Johan}} \yr{2023{\natexlab{{\em a\/}}}}
	\at{{An Improved Fringe-Region Technique for the Representation of Gravity
			Waves in Large Eddy Simulation with Application to Wind Farms}}.
	\jt{Boundary-Layer Meteorology}  \bvol{186}~(3),  \pg{567--593},
	\arxiv{arXiv: 2205.10612}.
	
	\bibitem[Lanzilao \& Meyers(2023{\natexlab{{\em
				b\/}}})]{lanzilao2023parametric}
	{\sc \au{Lanzilao, Luca} \& \au{Meyers, Johan}} \yr{2023{\natexlab{{\em b\/}}}}
	A parametric large-eddy simulation study of wind-farm blockage and gravity
	waves in conventionally neutral boundary layers,  \arxiv{arXiv: 2306.08633}.
	
	\bibitem[Luzzatto-Fegiz \& Caulfield(2018)]{Luzzatto-Fegiz2018}
	{\sc \au{Luzzatto-Fegiz, Paolo} \& \au{Caulfield, Colm Cille~P.}} \yr{2018}
	\at{{Entrainment model for fully-developed wind farms: Effects of atmospheric
			stability and an ideal limit for wind farm performance}}.  \jt{Physical
		Review Fluids}  \bvol{3}~(9),  \arxiv{arXiv: 1703.06553}.
	
	\bibitem[Maas(2022)]{Maas2022}
	{\sc \au{Maas, Oliver}} \yr{2022}  \at{{From gigawatt to multi-gigawatt wind
			farms: wake effects, energy budgets and inertial gravity waves investigated
			by large-eddy simulations}}.  \jt{Wind Energy Science Discussions}
	\bvol{2022}~(July),  \pg{1--31}.
	
	\bibitem[Maas(2023)]{Maas2023}
	{\sc \au{Maas, Oliver}} \yr{2023}  \at{{Large-eddy simulation of a 15 GW wind
			farm: Flow effects, energy budgets and comparison with wake models}}.
	\jt{Frontiers in Mechanical Engineering}  \bvol{9}~(March),  \pg{1--23}.
	
	\bibitem[Meyers {\em et~al.\/}(2022)Meyers, Bottasso, Dykes, Fleming, Gebraad,
	Giebel, G{\"{o}}{\c{c}}men \& van Wingerden]{Meyers2022}
	{\sc \au{Meyers, Johan}, \au{Bottasso, Carlo}, \au{Dykes, Katherine},
		\au{Fleming, Paul}, \au{Gebraad, Pieter}, \au{Giebel, Gregor},
		\au{G{\"{o}}{\c{c}}men, Tuhfe} \& \au{van Wingerden, Jan-Willem}} \yr{2022}
	\at{{Wind farm flow control: prospects and challenges}}.  \jt{Wind Energy
		Science}  \bvol{7}~(6),  \pg{2271--2306}.
	
	\bibitem[Niayifar \& Port{\'{e}}-Agel(2016)]{niayifar2016analytical}
	{\sc \au{Niayifar, Amin} \& \au{Port{\'{e}}-Agel, Fernando}} \yr{2016}
	\at{{Analytical modeling of wind farms: A new approach for power
			prediction}}.  \jt{Energies}  \bvol{9}~(9),  \pg{1--13}.
	
	\bibitem[Nishino \& Dunstan(2020)]{Nishino2020}
	{\sc \au{Nishino, Takafumi} \& \au{Dunstan, Thomas~D.}} \yr{2020}
	\at{{Two-scale momentum theory for time-dependent modelling of large wind
			farms}}.  \jt{Journal of Fluid Mechanics}  \bvol{894}.
	
	\bibitem[Patel {\em et~al.\/}(2021)Patel, Dunstan \& Nishino]{Patel2021}
	{\sc \au{Patel, Kelan}, \au{Dunstan, Thomas~D.} \& \au{Nishino, Takafumi}}
	\yr{2021}  \at{{Time-Dependent Upper Limits to the Performance of Large Wind
			Farms Due to Mesoscale Atmospheric Response}}.  \jt{Energies}
	\bvol{14}~(19),  \pg{6437}.
	
	\bibitem[Port{\'{e}}-Agel {\em et~al.\/}(2020)Port{\'{e}}-Agel, Bastankhah \&
	Shamsoddin]{PorteAgel2020}
	{\sc \au{Port{\'{e}}-Agel, Fernando}, \au{Bastankhah, Majid} \& \au{Shamsoddin,
			Sina}} \yr{2020}  \at{{Wind-Turbine and Wind-Farm Flows: A Review}}.
	\jt{Boundary-Layer Meteorology}  \bvol{174}~(1),  \pg{1--59}.
	
	\bibitem[Rampanelli \& Zardi(2004)]{Rampanelli2004}
	{\sc \au{Rampanelli, G.} \& \au{Zardi, Dino}} \yr{2004}  \at{{A method to
			determine the capping inversion of the convective boundary layer}}.
	\jt{Journal of Applied Meteorology}  \bvol{43}~(6),  \pg{925--933}.
	
	\bibitem[Sachsperger {\em et~al.\/}(2015)Sachsperger, Serafin \&
	Grubi{\v{s}}i{\'{c}}]{sachsperger2015lee}
	{\sc \au{Sachsperger, Johannes}, \au{Serafin, Stefano} \&
		\au{Grubi{\v{s}}i{\'{c}}, Vanda}} \yr{2015}  \at{{Lee waves on the
			boundary-layer inversion and their dependence on free-atmospheric
			stability}}.  \jt{Frontiers in Earth Sciences}  \bvol{3}~(November),
	\pg{1--11}.
	
	\bibitem[Smedman {\em et~al.\/}(1997)Smedman, Bergstr{\"{o}}m \&
	Grisogono]{Smedman1997}
	{\sc \au{Smedman, Ann-Sofi}, \au{Bergstr{\"{o}}m, Hans} \& \au{Grisogono,
			Branko}} \yr{1997}  \at{{Evolution of stable internal boundary layers over a
			cold sea}}.  \jt{Journal of Geophysical Research: Oceans}  \bvol{102}~(C1),
	\pg{1091--1099}.
	
	\bibitem[Smith(2007)]{Smith_2007}
	{\sc \au{Smith, Ronald~B.}} \yr{2007}  \at{{Interacting mountain waves and
			boundary layers}}.  \jt{Journal of the Atmospheric Sciences}  \bvol{64}~(2),
	\pg{594--607}.
	
	\bibitem[Smith(2010)]{Smith_2010}
	{\sc \au{Smith, Ronald~B.}} \yr{2010}  \at{{Gravity wave effects on wind farm
			efficiency}}.  \jt{Wind Energy}  \bvol{13}~(5),  \pg{449--458}.
	
	\bibitem[Stevens {\em et~al.\/}(2015)Stevens, Gayme \& Meneveau]{Stevens2015}
	{\sc \au{Stevens, Richard~J.A.M.}, \au{Gayme, Dennice~F.} \& \au{Meneveau,
			Charles}} \yr{2015}  \at{{Coupled wake boundary layer model of wind-farms}}.
	\jt{Journal of Renewable and Sustainable Energy}  \bvol{7}~(2),  \pg{1--25},
	\arxiv{arXiv: 1408.1730}.
	
	\bibitem[Stipa {\em et~al.\/}(2023{\natexlab{{\em a\/}}})Stipa, Ajay, Allaerts
	\& Brinkerhoff]{Stipa2023MSC}
	{\sc \au{Stipa, Sebastiano}, \au{Ajay, Arjun}, \au{Allaerts, Dries} \&
		\au{Brinkerhoff, Joshua}} \yr{2023{\natexlab{{\em a\/}}}}  \at{{The
			Multi-Scale Coupled Model : a New Framework Capturing Wind Farm-Atmosphere
			Interaction and Global Blockage Effects}}.  \jt{Wind Energy Science
		Discussions} ~(August),  \pg{1--44}.
	
	\bibitem[Stipa {\em et~al.\/}(2023{\natexlab{{\em b\/}}})Stipa, Ajay, Allaerts
	\& Brinkerhoff]{Stipa2023Tosca}
	{\sc \au{Stipa, S.}, \au{Ajay, A.}, \au{Allaerts, D.} \& \au{Brinkerhoff, J.}}
	\yr{2023{\natexlab{{\em b\/}}}}  \at{Tosca -- an open-source finite-volume
		les environment for wind farm flows}.  \jt{Wind Energy Science Discussions}
	\bvol{2023}~(May),  \pg{1--41}.
	
	\bibitem[Taylor \& Sarkar(2008)]{Taylor2008}
	{\sc \au{Taylor, John~R.} \& \au{Sarkar, Sutanu}} \yr{2008}
	\at{{Stratification effects in a bottom Ekman layer}}.  \jt{Journal of
		Physical Oceanography}  \bvol{38}~(11),  \pg{2535--2555}.
	
	\bibitem[Teixeira(2014)]{Orographic_gravity_wave_drag}
	{\sc \au{Teixeira, Miguel~A.C.}} \yr{2014}  \at{{The physics of orographic
			gravity wave drag}}.  \jt{Frontiers in Physics}  \bvol{2}~(July),
	\pg{1--24}.
	
	\bibitem[Troldborg \& {Meyer Forsting}(2017)]{Troldborg2017}
	{\sc \au{Troldborg, Niels} \& \au{{Meyer Forsting}, Alexander~R.}} \yr{2017}
	\at{{A simple model of the wind turbine induction zone derived from numerical
			simulations}}.  \jt{Wind Energy}  \bvol{20}~(12),  \pg{2011--2020}.
	
	\bibitem[Vosper(2004)]{vosper2004inversion}
	{\sc \au{Vosper, Simon~B.}} \yr{2004}  \at{{Inversion effects on mountain lee
			waves}}.  \jt{Quarterly Journal of the Royal Meteorological Society}
	\bvol{130}~(600 PART A),  \pg{1723--1748}.
	
	\bibitem[Zong \& Port{\'{e}}-Agel(2020)]{Zong2020}
	{\sc \au{Zong, Haohua} \& \au{Port{\'{e}}-Agel, Fernando}} \yr{2020}  \at{{A
			momentum-conserving wake superposition method for wind farm power
			prediction}}.  \jt{Journal of Fluid Mechanics}  \bvol{889}.
	
\end{thebibliography}


\end{document}